\documentclass[a4paper, 10pt]{article}
\usepackage[style=apa, 
backend=biber, 
uniquename=false,
url=true, 
doi=true]{biblatex}
\usepackage{hyperref}
\usepackage[dvipsnames]{xcolor}
\usepackage{colortbl}
\hypersetup{
plainpages=false,%
bookmarksopen=true,%
bookmarksnumbered=false,%
bookmarksdepth=5,
breaklinks=true,
colorlinks=true,
citecolor=BlueViolet,
linkcolor=BlueViolet,
urlcolor=blue,
}
\usepackage{citeauthor}
\usepackage{mathrsfs} 

\usepackage[T1]{fontenc}
\usepackage[utf8]{inputenc}

\usepackage{times} %change fonts only for texts with Nimbus Roman No.9

\usepackage{bm}
\usepackage{amsmath}
\usepackage{amssymb}

\usepackage{SpringerPsychometrika}

\usepackage{fancyhdr}

\fancypagestyle{firstpage}{
\fancyhf{}
\rhead{\textit{arXiv Preprint}}
\lhead{}
\chead{}
\cfoot{}
\lfoot{Published at \textit{Psychometrika} in 2022}

}

\fancypagestyle{fund_footer}{%
\fancyhf{}

\fancyfoot[L]{\small}
}
\usepackage{setspace}

\usepackage[noblocks, affil-it]{authblk}

\usepackage{abstract}
\renewenvironment{abstract}{\section*{\centering\abstractname}}{}
\renewcommand{\abstractname}{}

\usepackage{graphicx}
\newcommand{\orcid}[1]{\href{#1}{\includegraphics[width=10pt]{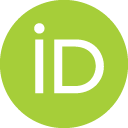}}}
\usepackage{graphicx}

\usepackage{booktabs}
\usepackage{multicol}
\usepackage{multirow}
\usepackage{dcolumn}
\usepackage{dcolumn}
\usepackage{lscape}

\newcolumntype{L}{>{$}l<{$}} 
\newcolumntype{C}{>{$}c<{$}} 
\newcolumntype{R}{>{$}r<{$}} 

\usepackage{float}

\usepackage{threeparttable}
\usepackage{adjustbox}
\usepackage{algorithm}
\usepackage{algorithmic}
\newcommand{\argmax}{\mathop{\rm arg~max}\limits}

\begin{document}

% Title
\title{\vspace{-1.5\bigskipamount} \normalsize \uppercase{Scalable Bayesian Approach For The Dina Q-Matrix Estimation Combining Stochastic Optimization And Variational Inference}}
\author[ ]{\fontsize{11pt}{11pt}{\scshape Motonori Oka}}
\author[ ]{\fontsize{11pt}{11pt}{\scshape Kensuke Okada}}
\affil[ ]{\small{GRADUATE SCHOOL OF EDUCATION, THE UNIVERSITY OF TOKYO}}
\date{}
\maketitle

\thispagestyle{firstpage}

\vspace{-2.7\bigskipamount}
\begin{abstract}
\begin{center}\footnotesize
\begin{minipage}{\dimexpr\paperwidth-79mm}
\setlength{\parindent}{12pt}
Diagnostic classification models (DCMs) offer statistical tools to inspect the fined-grained attribute of respondents' strengths and weaknesses. \textcolor{black}{However, the diagnosis accuracy deteriorates when misspecification occurs in the predefined item-attribute relationship, which is encoded into a Q-matrix. To prevent such misspecification, methodologists have recently developed several Bayesian Q-matrix estimation methods for greater estimation flexibility. However, these methods become infeasible in the case of large-scale assessments with a large number of attributes and items}. In this study, we focused on the deterministic inputs, noisy ``and'' gate (DINA) model and proposed a new framework for the Q-matrix estimation to find the Q-matrix with the maximum marginal likelihood. Based on this framework, we developed a scalable estimation algorithm for the DINA Q-matrix by constructing an iteration algorithm that utilizes stochastic optimization and variational inference. 
The simulation and empirical studies reveal that the proposed method achieves high-speed computation, good accuracy\textcolor{black}{, and robustness to potential misspecifications, such as initial value choices and hyperparameter settings.} Thus, the proposed method can be a useful tool for estimating a Q-matrix in large-scale settings.\\

\end{minipage}
\end{center}
\end{abstract}

\vspace{0.9\bigskipamount}
\section{Introduction}
One of the recent advances in psychometrics concerns the class of statistical models known as diagnostic classification models (DCMs), which offer a tool for measuring examinees' strengths and weaknesses in their latent states. In DCMs, these states are often assumed to comprise a set of binary skills, which are referred to as attributes. The estimated mastery or non-mastery of the attributes constitutes the attribute mastery profile of the examinee. By providing detailed information on the examinees' strengths and weaknesses, which are reflected in their attribute mastery profiles, DCMs permit teachers to engage in classroom instruction and learning with a focus on the students' educational needs. Various DCMs have been developed to capture diverse response-generating processes for accurate diagnosis \parencite[for example,][]{delatorre_generalized_2011, hartz_fusion_2008, henson_defining_2009, von_davier_general_2008}. In this study, we focus on the deterministic inputs, noisy ``and'' gate \parencite[DINA;][]{junker_cognitive_2001} model. This model essentially assumes that an examinee must possess all the attributes required by an item to answer it correctly. Owing to its parsimony and interpretability, the DINA model has been widely used in both methodological and applied research in educational and psychological assessments.

Although many other sub-models have been proposed, essentially, all DCMs share one common characteristic: the item-attribute relationship is embodied in a Q-matrix \parencite{tatsuoka_rule_1983}. Its binary element is denoted by $q_{jk} \in \{0,1\}$, where $j=1,\ldots,J$ and $k=1,\ldots,K$ correspond to the item and attribute indices. Statistical analysis based on DCMs requires researchers to specify a Q-matrix. If $q_{jk}$ equals one, then the correct response for item $j$ necessitates attribute $k$ and vice versa. The elements of a Q-matrix determine whether the item requires the mastery of specific attributes, and a Q-matrix is often constructed based on the knowledge of domain experts. However, in practice, Q-matrix misspecification occurs frequently because of, for example, the inadequacy of the underlying theories. Such misspecification leads to a deterioration in the accuracy of parameter estimation and diagnosis \parencite{rupp_effects_2008,kunina-habenicht_impact_2012}. To identify the misfit of a Q-matrix to data, researchers have recently developed several Q-matrix estimation algorithms using different approaches \parencite{chen_statistical_2015, chen_bayesian_2018, chung_gibbs_2019, culpepper_estimating_2019, decarlo_recognizing_2012, liu_data-driven_2012, liu_theory_2013, xu_identifying_2018,liu_constrained_2020, chen_sparse_2020, culpepper_development_2019, culpepper_exploratory_2019, gu_joint_2021}. This study considers the Q-matrix estimation to binary response data from a Bayesian paradigm.

Previous studies on Bayesian Q-matrix estimation rely on Markov chain Monte Carlo (MCMC) methods. The mainstream of these studies can be grouped into two major categories: those that estimate some elements of a Q-matrix and those that estimate the entire Q-matrix. The former type regards a large part of a Q-matrix as known and infers the posterior of the remaining unknown elements, which are assumed to follow Bernoulli distributions, with Beta priors \parencite{decarlo_recognizing_2012}. This method is computationally less intensive. However, a critical problem is that the plausibility of the prespecified Q-matrix elements that are assumed to be known cannot be assessed.

For the latter type, several previous studies have proposed MCMC methods for estimating the entire Q-matrix. \textcite{chen_bayesian_2018} addressed the problem of identifiability in estimating a Q-matrix by developing the constrained Gibbs sampler that restricts the sampling space to identified Q-matrices. Similarly, \textcite{liu_constrained_2020} developed the Q-matrix estimation algorithm by combining the Robbins-Monro scheme and the MCMC method with the identification constraints on a Q-matrix. Their algorithm achieved faster convergence in estimating the parameters for the DINA model than the constrained Gibbs sampler \parencite{chen_bayesian_2018}. Owing to its better convergence property, it enjoys feasible computation for estimating the identified Q-matrix with a large number of attributes such as $K=9$. These two algorithms expanded the applicability of Q-matrix estimation in a fundamental manner. \textcite{culpepper_estimating_2019} proposed another estimation method that incorporates domain-expert knowledge into prior distribution. This strategy facilitates the interpretability of an estimated Q-matrix and helps identify residual attributes that are not within the scope of prior knowledge. This method also enables explainable reasoning for the Q-matrix estimation process and offers a means of validating the underlying theory that is utilized in constructing a Q-matrix. \textcite{chung_gibbs_2019} developed a generic Gibbs sampler that does not necessitate any constraints or prior specifications. Based on the Monte Carlo simulation, the results of this study revealed that, even in the condition of an incomplete and complex true Q-matrix with the small size of items, this generic Gibbs sampler could estimate approximately $85 \%$ entries of such a Q-matrix. Other estimation methods have also been developed to expand the utility and applicability of the Q-matrix estimation, such as estimating a Q-matrix under a more flexible DCM \parencite{culpepper_exploratory_2019} and inferring both Q-matrix's structures and types of item-responding processes behind test items \parencite{chen_sparse_2020}. In the case of an assessment with relatively small sizes of attributes and items, these MCMC-based Q-matrix estimation methods can provide a reliable Q-matrix estimate in a feasible time. However, a large-scale assessment with high-dimensional attributes implies an unaffordable computational load because the parameter space of a Q-matrix expands exponentially with an increase in the number of attributes. This infeasibility of the Q-matrix estimation in large-scale settings is a common problem in MCMC-based Q-matrix estimation methods.

Although \textcite{liu_constrained_2020} made an important stride in estimating the identified Q-matrix with computational efficiency and showed the sound feasibility of their algorithm when estimating a Q-matrix with $K=9$ attributes, its accuracy was not satisfactory under the condition where no informative starting values were provided because of the vast sampling space of identified Q-matrices. In addition, enforcing the identification constraints on a Q-matrix will be practically difficult in the case of a large $K$ due to the increase in the number of items to satisfy the identification conditions. For example, the mathematics dataset of the Trends in International Mathematics and Science Study (TIMSS) 2003, which is one of the frequently-used real datasets in the DCM's literature, has 23 items pertaining to 13 attributes. To assess the possibility of this size of a Q-matrix to hold an identified structure, we randomly generated 1 billion Q-matrices with 23 items and 13 attributes and did not find the one with such a structure in the generated Q-matrices. This indicates that estimating a Q-matrix with the identification constraints would fail under a large number of attributes and a relatively small size of items. Moreover, the parameter space of Q-matrices without these constraints is significantly larger than that of Q-matrices with these constraints, leading to a stricter computational requirement of the Q-matrix estimation. Specifically, while the generic Gibbs sampler by \textcite{chung_gibbs_2019} requires three chains of 100,000 iterations with 50,000 burn-in for convergence, the constrained Gibbs sampler by \textcite{chen_bayesian_2018} requires one chain of 30,000 iterations with 15,000 burn-in. This indicates that the Q-matrix estimation without identification constraints is much more computationally intensive than with these constraints. Therefore, a scalable generic Bayesian Q-matrix estimation algorithm applicable to large-scale assessments should be developed. 

To address the problem of scalability in the Bayesian Q-matrix estimation, we consider a novel problem setting for estimating a Q-matrix. Specifically, we regard the Q-matrix estimation as a root-finding problem with the objective of searching for the Q-matrix that produces the \emph{maximum marginal likelihood}. The marginal likelihood $p(\mathbf{X} \vert M)= \int p(\mathbf{X} \vert \bm{\Theta}_M, M)d\bm{\Theta}_M$, which is also known as \emph{model evidence}, marginalizes out model parameters $\bm{\Theta}_M$ and quantifies the probability of observing the data $\mathbf{X}$ given the assumption of model $M$ \parencite{lee_bayesian_2013}. $\bm{\Theta}_M$ denotes model parameters pertaining to model $M$. In terms of DCMs, a model assumption $M$ encompasses the DCM and Q-matrix specifications. As a Q-matrix can be regarded as a specification of a model that determines item-attribute relationships, it is theoretically justifiable to select the Q-matrix with the maximum marginal likelihood as the optimal solution in the same manner as model selection based on model fit measures. However, the parameter space of a Q-matrix generally consists of $2^{JK}$ possible Q-matrices. Therefore, the optimal solution among them is challenging to pinpoint. For this problem, we propose a novel iteration algorithm for Q-matrix optimization that utilizes {\it stochastic optimization} and {\it variational inference}. Both techniques reduce computational complexity, and a number of recent developments in machine learning make use of them owing to their greater scalability \parencite[for example,][]{blei_latent_2003, hoffman_stochastic_2013}. Regarding the former technique, we \textcolor{black}{approximate the log marginal likelihood with mini-batch samples} to alleviate intensive computation. Regarding the latter, we employ the newly developed variational Bayes algorithm for parameter estimation of the DINA model \parencite{yamaguchi_variational_2020}.

The remainder of this paper is organized as follows. In Section 2, we first introduce the DINA model tailored for variational inference. Second, we formulate the problem setting for the Q-matrix estimation. Third, the two major techniques that constitute the proposed method, stochastic optimization, and variational inference, are presented. In Section 3, we report the results of a simulation study that confirms how accurately the proposed method estimates the Q-matrix under various conditions. Then, in Section 4, we report empirical studies that used three real datasets to compare the computation time among the proposed method, Gibbs sampler \parencite{chung_gibbs_2019}, and EM-based algorithm with Lasso regularization \parencite{chen_statistical_2015}. Additionally, we comparatively evaluate the relative model fit. In Section 5, we discuss the effects of potential misspecifications on the Q-matrix estimation, such as initial value choices and hyperparameter settings. Finally, in Section 6, we summarize the important findings from the simulation and empirical studies and discuss the limitations and future directions of this study.

\section{Method}
\subsection{Formulation of the DINA Model}
The following indices $i\; (1,\ldots,N)$, $j\; (1,\ldots,J)$, and $k\; (1, \ldots, K)$ are used in this article to denote respondents, items, and attributes, respectively. The number of latent classes defined by attribute mastery profiles amounts to $2^K$, and each latent class is indexed by $l\; (1,\ldots,2^K=L)$. Let $\bm{\alpha}_l=(\alpha_{l1}, \ldots, \alpha_{lk}, \ldots, \alpha_{lK})^{\top}$ and $\bm{q}_j=(q_{j1}, \ldots, q_{jk}, \ldots, q_{jK})^{\top}$ be the $l$-th row vector of an attribute mastery profile pattern matrix $\mathbf{A}=(\bm{\alpha}_1, \ldots, \bm{\alpha}_l, \ldots, \bm{\alpha}_L)^{\top}$ and the $j$-th row vector of a Q-matrix $\mathbf{Q}=(\bm{q}_1, \ldots, \bm{q}_j, \ldots, \bm{q}_J)^{\top}$, where the transpose is denoted by the superscript $\top$. In this paper, we call the row vectors of a Q-matrix \emph{$q$-vectors}, and each $q$-vector $\bm{q}_j$ represents a Q-matrix specification for item $j$.
Additionally, we designate $\mathbf{X}=(\bm{x}_1, \ldots, \bm{x}_j, \ldots, \bm{x}_J)^\top$ as the $N \times J$ response matrix whose elements are binary random variables $x_{ij}$.
\textcite{yamaguchi_variational_2020} reformulated the DINA model as a mixture model by introducing the binary latent class indicator vector $\bm{z}_i=(z_{i1},\ldots, z_{il}, \ldots, z_{iL})^{\top}$. Its element $z_{il}$ takes the value of 1 if a respondent possesses a set of attributes corresponding to class $l$, and 0 otherwise. Hence, the elements of $\bm{z}_i$ satisfy $z_{il} \in \{0,1\}$ and $\sum_{l=1}^L z_{il}=1$. As an example, let us consider the case of $K=2$ and $L=2^2=4$. A respondent who has mastered none of the required attributes belongs to class 1 and is given the latent class indicator vector $\bm{z}_i=(1,0,0,0)^{\top}$. Similarly, a respondent who has acquired all the attributes is classified into class 4 and is given $\bm{z}_i=(0,0,0,1)^{\top}$. Respondents in the other two classes are also classified according to $\bm{z}_i$.

By incorporating the latent variable $\bm{z}_i$, the ideal response for each respondent can be expressed as $\eta_{ij}=\bm{\eta}_j^{\top} \bm{z}_i=\sum_{l=1}^L \eta_{lj} z_{il}$, where $\eta_{lj}= \prod_{k=1}^K \alpha_{lk}^{q_{jk}}$ is the ideal response to item $j$ for a respondent in class $l$. Because the DINA model is noncompensatory, $\eta_{lj}$ equals 1 only when the respondents in class $l$ have mastered all the attributes required by item $j$. The vectorized form of ideal responses is $\bm{\eta}_j=(\eta_{1j}, \ldots, \eta_{lj}, \ldots, \eta_{Lj})^{\top}$. 

Finally, with the introduction of the slip $s_j=p(x_{ij}=0 \vert\eta_{ij}=1)$ and guessing $g_j=p(x_{ij}=1\vert \eta_{ij}=0)$ parameters, the item response function of the DINA model is given as
\begin{align*}
p(x_{ij} \vert s_j, g_j,\bm{z}_i,\bm{q}_j,\bm{\alpha}_l)= \Bigl( (1-s_j)^{x_{ij}} s_j^{1-x_{ij}} \Bigl)^{\bm{\eta}_j^{\top}\bm{z}_i} \Bigl(g_j^{x_{ij}}(1-g_j)^{1-x_{ij}} \Bigl)^{1-\bm{\eta}_j^{\top} \bm{z}_i}.
\end{align*}
The slip $s_j$ parameter represents the probability of obtaining an incorrect response for item $j$ when a respondent masters all the required attributes, and the guessing parameter $g_j$ represents the probability of obtaining a correct response when a respondent lacks at least one of the required attributes. This formulation paves the way to well-studied estimation methods, such as expectation-maximization (EM) and variational Bayes (VB) algorithms for mixture models \parencite{bishop_pattern_2006}, and enables the derivation for estimating the parameters of interest to be more tractable. 

\subsection{Problem Setting}
We frame Q-matrix estimation as a root-finding problem, where the objective function is set to the log marginal likelihood $\log p(\mathbf{X}\vert \mathbf{Q})$. This notation indicates that a Q-matrix $\mathbf{Q}$ is the focus of interest among the specifications of model $M$. The goal of a root-finding problem is to locate a point $\theta^*$ that satisfies the equation $f(\theta^*)=0$ for a given function $f(\theta)$. In this case, the proposed algorithm aims to obtain at least one root $\mathbf{Q}^* \in \mathscr{Q} =\{ \mathbf{Q}_1,\ldots,\mathbf{Q}_{2^{JK}}\}$ to $\log p(\mathbf{X}\vert \mathbf{Q})- \log p(\mathbf{X}\vert \mathbf{Q}^*)=0$. Here, $\mathscr{Q}$ is the Q-matrix space comprising $2^{JK}$ possible Q-matrices, and $\mathbf{Q}^*$ is the Q-matrix that yields the maximum log marginal likelihood. Note that, owing to the monotonicity of the logarithmic function, the Q-matrix with the maximum log marginal likelihood also maximizes the marginal likelihood.

The general approach to a root-finding problem is based on an iteration algorithm that produces a sequence of updated values for a target variable, and the updated values in that sequence are expected to converge toward the root at a limit. In practice, however, we cannot proceed with the iteration until the limit because of finite time and computational resources, so that we need to end the iteration at an appropriate timing. Hence, the final value of the iteration becomes an approximation of the root. The accuracy of this approximation often improves as the number of iterations increases, and we can stop the algorithm after achieving the desired level of accuracy. In this study, we formulate the Q-matrix estimation as a root-finding problem and then develop an iteration algorithm to find the Q-matrix that optimizes the log marginal likelihood.

\subsection{Proposed Method}
Two problems emerge when constructing the iteration algorithm for Q-matrix estimation. First, it is computationally expensive to compute the log marginal likelihood given all data points when evaluating the goodness of each updated value at every iteration. Second, the log marginal likelihood $\log p(\mathbf{X}\vert \mathbf{Q})$ itself is intractable. The proposed algorithm resolves the first problem through \textcolor{black}{approximating $\log p(\mathbf{X}\vert \mathbf{Q})$ with mini-batch samples}, which alleviates the intensive computation of the log marginal likelihood. For the second problem, we replace the log marginal likelihood with its variational lower bound (ELBO) as a proxy. These two approximation techniques enable a scalable estimation algorithm for a Q-matrix. We first present the concept of stochastic optimization in Section 2.3.1, and then explain the idea of variational Bayes and the variational approximation to the log marginal likelihood in Section 2.3.2. In Section 2.3.3, we elaborate on the recursive update rule for a Q-matrix using posterior probabilities of possible Q-matrices based on those of possible $q$-vectors for items. The entire picture of the proposed algorithm is presented later as pseudocode. 

\subsubsection{Stochastic Optimization}
Stochastic optimization methods aim to solve a stochastic root-finding problem in which only the noisy measurements of the true objective function are observable \parencite{spall_introduction_2003}. The cornerstone of this field was laid by \textcite{robbins_stochastic_1951}, who proposed a recursive update scheme to find the root of a function. This scheme has served as the theoretical foundation for crucial optimization techniques such as the stochastic gradient descent algorithm in the modern machine learning literature \parencite{mandt_stochastic_2017}.

The basic concept of stochastic optimization is to find the root $\theta^*$ of an objective function $g(\theta)$ through the approximation of that function $f(\theta, W)$, where $W$ represents a random variable utilized to approximate $g(\theta)$. This approximation is conducted stochastically with the guarantee that $f(\theta, W)$ is an unbiased estimator of $g(\theta): g(\theta)=\mathrm{E}_W [f(\theta,W)]$. Reasons for introducing \textcolor{black}{the objective function approximated with random variables} include, for example, the fact that its true function $g(\theta)$ is not known, intractable, or computationally intensive. 

\textcolor{black}{These types of methods} also have been used in psychometrics, especially for performing estimation with an EM-like algorithm for a model in which \textcolor{black}{the expected complete-data log-likelihood function} cannot be evaluated in closed form. In this case, some function for parameter estimation, such as \textcolor{black}{the expected complete-data gradient with respect to model parameters}, is approximated with Monte Carlo samples of missing data \parencite{delyon_convergence_1999}. The relevant and widely used method in this discipline is the Metropolis-Hastings Robbins-Monro (MH-RM) algorithm \parencite{cai_high-dimensional_2010, cai_metropolis-hastings_2010}. Several psychometric methods have applied this algorithm for their estimation, such as confirmatory item factor analysis \parencite{cai_metropolis-hastings_2010}, noncompensatory multidimensional item response theory (IRT) models \parencite{chalmers_maximum-likelihood_2014}, and Q-matrix estimation \parencite{liu_constrained_2020}. The MH-RM algorithm replaces the E step in the conventional EM algorithm \parencite{dempster_maximum_1977} with a stochastic imputation step to obtain Monte Carlo samples of missing data including respondent parameters. \textcolor{black}{Subsequently, to circumvent the numerical integration in evaluating the expected complete-data gradient and information matrix, the gradient is approximated with simulated Monte Carlo samples, and the information matrix is updated with a recursive approximation procedure of the expected complete-data information matrix using simulated Monte Carlo samples. Lastly, parameter values are updated iteratively under a Robbins-Monro scheme with the approximated gradient and information matrix. When Monte Carlo samples are simulated from their exact posteriors in such cases as DCMs, the approximation of the expected complete-data gradient becomes its unbiased estimator, otherwise it is approximately unbiased \parencite{zhang_computation_2022}.} Other uses of this type of methods can be found in latent regression models \parencite{von_davier_stochastic_2010} and exploratory IRT models \parencite{camilli_stochastic_2019}.

In the proposed method, our objective function becomes \textcolor{black}{the log marginal likelihood $\log p(\mathbf{X}_S \vert \mathbf{Q})$ approximated with mini-batch samples}, rather than $\log p(\mathbf{X}\vert \mathbf{Q})$, where $\mathbf{X}_S$ is a randomly selected subset comprising $S$ samples \textcolor{black}{(mini-batch samples)} from the dataset $\mathbf{X}$. The log marginal likelihood $\log p(\mathbf{X} \vert \mathbf{Q})=\sum_{j=1}^J\sum_{i=1}^N \log p(x_{ij} \vert \mathbf{Q})$ fulfills the following equivalence in the same manner as \textcolor{black}{approximating the log likelihood with mini-batch samples} \parencite{naesseth_machine_2018}:
\begin{align*}
\sum_{j=1}^J\sum_{i=1}^N \log p(x_{ij} \vert \mathbf{Q}) &= \sum_{j=1}^J\sum_{s=1}^S \mathrm{E}\left[ \frac{N}{S}\log p(x_{\tau_s j} \vert \mathbf{Q})\right] \\
& (j=1,\ldots,J;\; i=1,\ldots,N;\; s=1,\ldots,S),
\end{align*}
where indices $\bm{\tau}=\{\tau_1, \ldots, \tau_S \}$ are sampled from the discrete uniform distribution $\mathrm{DiscreteUniform}(1,N)$. By using \textcolor{black}{the mini-batch samples} from all data points at every iteration, we attain high scalability toward a large-scale setting. After the stochastic optimization, the problem setting is rewritten to obtain at least one root $\mathbf{Q}^* \in \mathscr{Q} =\{ \mathbf{Q}_1,\ldots,\mathbf{Q}_{2^{JK}}\}$ that satisfies $\mathrm{E}[\frac{N}{S} \log p(\mathbf{X}_S \vert \mathbf{Q})] - \log p(\mathbf{X}\vert \mathbf{Q}^*)=0$. 

Beyond the computational advantage, we observed that the stochastic variability emerging from shuffling \textcolor{black}{the mini-batch samples} $\mathbf{X}_S$ during iteration causes a significant increase in the effectiveness of the optimization process. In Section 3.2, we show how stochastic variability contributes to an effective search for the optimal Q-matrix. For a more comprehensive review and mathematical treatment of stochastic optimization, refer to \textcite{spall_introduction_2003}.

\subsubsection{Variational Inference}
\subsubsection*{Variational Bayes (VB)}
In this section, we introduce the general concept of variational inference. The main objective in Bayesian statistics is to infer the posterior distributions of the latent variables $\bm{\Theta}$ \parencite{gelman_bayesian_2013}, and this inference follows the consistent principle: Bayes rule
\begin{align*}
p(\bm{\Theta}\vert \mathbf{X}) = \frac{p(\mathbf{X}\vert \bm{\Theta})p(\bm{\Theta})}{\int p(\mathbf{X }\vert\bm{\Theta})p(\bm{\Theta})d\bm{\Theta}}.
\end{align*}
However, in most applications, the integral in the denominator is in high dimension. Thus, the computation of this normalizing constant is often not tractable.

Variational inference circumvents this intractability by reframing the posterior computation as an optimization problem, with the goal of minimizing the divergence measure between the variational distribution $q(\bm{\Theta})$ and the posterior distribution $p(\bm{\Theta}\vert \mathbf{X})$ \parencite{blei_variational_2017}. The variational distribution is introduced to approximate the posterior distribution, and the divergence measure between them quantifies the degree of how well $q(\bm{\Theta})$ approximates $p(\bm{\Theta}\vert \mathbf{X})$.

The divergence emerges from deriving the lower bound of the log marginal likelihood $\log p(\mathbf{X})$. By applying Jensen's inequality, $\log p(\mathbf{X})$ can be decomposed as
\begin{align*}
\log p(\mathbf{X}) &= \int q(\bm{\Theta})\log\frac{p(\mathbf{X},\bm{\Theta})}{q(\bm{\Theta})}d\bm{\Theta} - \int q(\bm{\Theta})\log\frac{p(\bm{\Theta} \vert \mathbf{X})}{q(\bm{\Theta})}d\bm{\Theta} \\
&= L(q, \mathbf{X}) + \mathrm{KL}[ q(\bm{\Theta})\parallel p(\bm{\Theta}\vert \mathbf{X})]\\
&\geq L(q, \mathbf{X}),
\end{align*}
where $L(q, \mathbf{X})$ is the lower bound of the log marginal likelihood, which is also known as the evidence lower bound (ELBO), and KL$[q(\bm{\Theta}) \parallel p(\bm{\Theta}\vert \mathbf{X})]$ is the Kullback-Leibler (KL) divergence between the variational distribution $q(\bm{\Theta})$ and the posterior $p(\bm{\Theta}\vert \mathbf{X})$ \parencite{beal_variational_2003}. Reducing the KL divergence to 0 or equivalently maximizing the ELBO to $\log p(\mathbf{X})$ results in the equivalence between $q(\bm{\Theta})$ and $p(\bm{\Theta}\vert \mathbf{X})$. In practice, we resort to the mean-field assumption on $q(\bm{\Theta})$ for ease of optimization. This implies that $q(\bm{\Theta})$ is assumed to be factorized as
\begin{align*}
q(\bm{\Theta}) = \prod_{g=1}^G q(\bm{\Theta}_g).
\end{align*}
Each set of parameters for $q(\bm{\Theta}_g)$ is selected to satisfy $q(\bm{\Theta}_g) \propto \exp(\mathrm{E}_{g\neq h}[ \log p(\mathbf{X}, \bm{\Theta})])$, and the optimization of $q(\bm{\Theta}_g)$ continues until the value of the ELBO converges. The final $q(\bm{\Theta}_g)$ is known as the variational posterior for $\bm{\Theta}_g$. Here, $G$ denotes the number of parameters to be estimated, and $E[\cdot]_{g \neq h}$ denotes the expectation of variational posteriors with respect to all the parameters, except for the $g$-th parameter. For instance, $G$ equals $N\times 2^K + 2^K + 2J$ for the DINA model, where each term corresponds to the number of latent class indicator variables, structural parameters, and item parameters, respectively.

As the ELBO is the lower bound of the log marginal likelihood, it is a measure of how well the model fits the data. Thus, the ELBO can be used for model selection \parencite{zhang_advances_2019}. With the power of deterministic approximation of the posterior distribution, variational inference enables fast computation and exhibits high scalability to large datasets. These characteristics are in contrast to those of MCMC-based methods.

In the proposed method, we do not consider a Q-matrix as one of the model parameters. Rather, a Q-matrix is regarded as a key component of the model structure. As mentioned in the Introduction, our framework aims to find the Q-matrix, or more specifically the model with the specified Q-matrix structure, that yields the maximum marginal likelihood. This treatment of a Q-matrix as a structural model component is in line with typical methodological and empirical studies in DCMs, in which a Q-matrix is prespecified in the data analysis. With such typical context in DCMs, our model parameters $\bm{\Theta}$ are respondents' attribute mastery profiles (which are obtained from latent class indicator variables) and \textcolor{black}{structural/}item parameters. These model parameters are estimated by variational inference given the model that has a certain specification of a Q-matrix. Thus, the log marginal likelihood and its decomposition are rewritten such that
\begin{align*}
\log p(\mathbf{X}\vert \mathbf{Q}) &= \int q(\bm{\Theta}\vert \mathbf{Q})\log\frac{p(\mathbf{X},\bm{\Theta}\vert \mathbf{Q})}{q(\bm{\Theta}\vert \mathbf{Q})}d\bm{\Theta} - \int q(\bm{\Theta}\vert \mathbf{Q})\log\frac{p(\bm{\Theta} \vert \mathbf{X}, \mathbf{Q})}{q(\bm{\Theta}\vert \mathbf{Q})}d\bm{\Theta} \\
&\geq L(q, \mathbf{X} \vert \mathbf{Q}).
\end{align*}
Then, we aim to estimate the DINA model parameters with the certain specification of a Q-matrix and use these estimates for selecting an updated Q-matrix for each iteration, which we introduce in the following sections.

Variational Bayes has also been employed in estimation methods for latent variable models in psychometrics to increase the scalability of their Bayesian estimation. In the DCM's literature, VB methods have been developed for several models, such as the DINA model \parencite{yamaguchi_variational_2020}, saturated DCM \parencite{yamaguchi_variational_2021}, and multiple-choice DINA model \parencite{yamaguchi_vbmultiplechoice_2020}. All the three methods reformulate these DCMs into the form of mixture models to obtain conditional conjugacy for better mathematical and computational tractability of VB inference. Their studies shed light on the desirable property of VB methods. For instance, a study on the DINA model \parencite{yamaguchi_variational_2020} revealed that their VB method could stably estimate item parameters under small-scale conditions where the ML method produced the irregular estimates of these parameters. Further, a study on the saturated DCM \parencite{yamaguchi_variational_2021} also showed that their VB method exhibited better computational efficiency than the ML method under conditions that resemble computerized adaptive testing. In the IRT's literature, \textcite{natesan_bayesian_2016} employed a local variational method that introduces the bound of the logistic function to enable the tractable derivation of a VB algorithm for IRT models. \textcite{rijmen_fitting_2013} applied VB inference to estimate IRT models with random item effects for large-scale educational assessments. Moreover, to address the computational burden on estimating a multidimensional IRT, \textcite{cho_gaussian_2021} developed its VB algorithm that scales to high-dimensional settings. For other models, \textcite{humphreys_variational_2003} utilized a VB method to alleviate intensive computation for a latent class model. \textcite{jeon_variational_2017} proposed a VB algorithm for generalized linear mixed models with crossed random effects. As stated above, many researchers have recently resorted to VB inference for scalable Bayesian estimation under large-scale conditions.

\subsubsection*{Variational approximation to the log marginal likelihood}
\begin{figure}[!htpb]
 \begin{center}
 \centering
 \includegraphics[width=\linewidth]{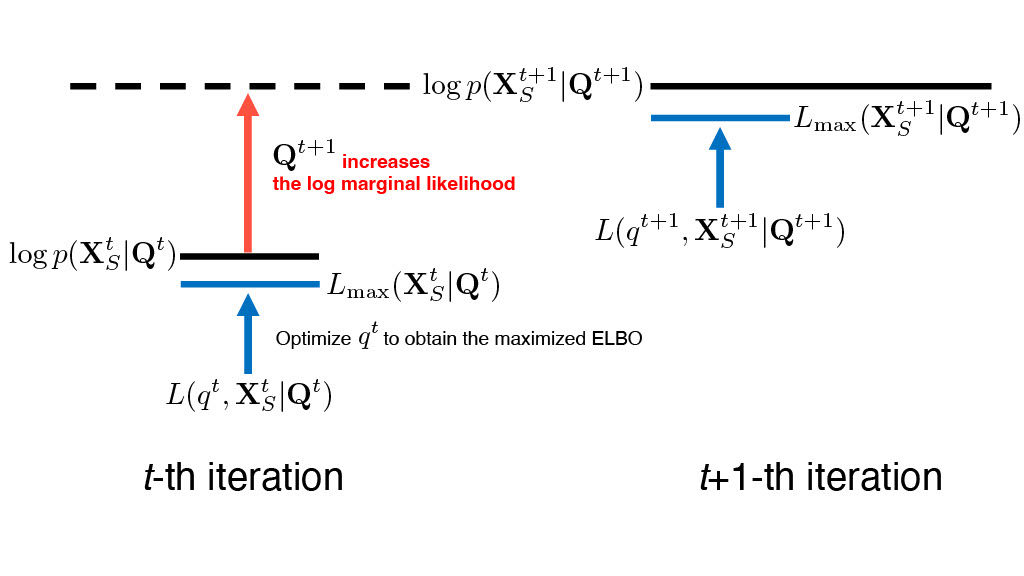}
\caption{A schematic of the typical transition of the maximized ELBO during iteration. The notation $q^t$ in the ELBO $L(q^t, \mathbf{X}_S^{t}\vert \mathbf{Q}^t)$ represents the variational distribution $q^t(\bm{\Theta} \vert \mathbf{X}_S^t, \mathbf{Q}^t)$ at the $t$-th iteration. In each iteration, we optimize $q^t(\bm{\Theta} \vert \mathbf{X}_S^t, \mathbf{Q}^t)$ to obtain the maximized ELBO $L_{\mathrm{max}} (\mathbf{X}_S^{t} \vert \mathbf{Q}^t)$ (blue arrow), which approximates the log marginal likelihood $\log p(\mathbf{X}_S^{t} \vert \mathbf{Q}^t)$ (black line on the left). The updated Q-matrix in the next $t+1$-th iteration, $\mathbf{Q}^{t+1}$, is expected to increase the log marginal likelihood (black line on the right) so that the maximized ELBO at $t+1$ also produces a larger value of the ELBO than the one at $t$.}
\label{fig:transition_SAELBO}
 \end{center}
\end{figure}

The second problem of forming the iteration algorithm for Q-matrix estimation is that the log marginal likelihood is not tractable. To tackle this intractability, we use the ELBO as a surrogate of the log marginal likelihood. Then, the stochastic objective function $\log p(\mathbf{X}_S \vert \mathbf{Q})$ in the current problem setting can be replaced with the maximized ELBO $L_{\mathrm{max}}(\mathbf{X}_S \vert \mathbf{Q})$, where the two quantities satisfy
\begin{align*}
\log p(\mathbf{X}_S \vert \mathbf{Q}) \geq L_{\mathrm{max}}(\mathbf{X}_S \vert \mathbf{Q}).
\end{align*}
To obtain the maximized ELBO given $\mathbf{Q}$, we optimize the variational distribution $q(\bm{\Theta}\vert \mathbf{Q})$ in the $L(q, \mathbf{X}_S \vert \mathbf{Q})$ at every iteration to approximate the corresponding posterior. Note that this optimization is for $q(\bm{\Theta} \vert \mathbf{Q})$ and not for $\mathbf{Q}$. In this sense, $q(\bm{\Theta}\vert \mathbf{Q})$ is optimized \emph{within each iteration} to obtain the maximized ELBO $L_{\mathrm{max}}(\mathbf{X}_S \vert \mathbf{Q})$ given the current specification of $\mathbf{Q}$. Hence, $q(\bm{\Theta} \vert \mathbf{Q})$ takes different parametric forms depending on different specifications of $\mathbf{Q}$. The iteration algorithm for a Q-matrix then pursues the Q-matrix with the globally maximized ELBO $L_{\mathrm{max}}(\mathbf{X}_S \vert \mathbf{Q}^*)$. A graphical illustration of the transition of the maximized ELBO is shown in Figure \ref{fig:transition_SAELBO}. 

Using the two types of approximations to the objective function, we are now able to develop an iteration algorithm for the Q-matrix estimation.

\subsubsection{Recursive Update Rule of a Q-matrix}
By stochastic optimization and variational approximation to the log marginal likelihood, the objective function becomes tractable. Based on this, we develop a specific method for updating a Q-matrix to pinpoint the optimal Q-matrix among $2^{JK}$ possible Q-matrices. As shown in Figure \ref{fig:transition_SAELBO}, an updated Q-matrix is expected to produce a larger value of the maximized ELBO. To construct the mechanism to find such a Q-matrix in each iteration, we employed the concept of the likelihood and posterior probability of \emph{possible $q$-vector patterns} for items, which was originally developed by \textcite{chung_gibbs_2019}. These quantities enable us to define the optimality of the updated Q-matrices. That is, we select the updated Q-matrix as the one with the highest posterior probability among the $(2^K-1)^J$ possible Q-matrices. Then, variational Bayes estimates, which are obtained from the process in the variational approximation to the log marginal likelihood, are incorporated as the components comprising the posterior probability of possible $q$-vector patterns for items. These estimates deliver information about the goodness of $\mathbf{Q}$ from one iteration to the next.

In the subsequent parts of this section, we first introduce the idea of the likelihood and posterior probability of possible $q$-vector patterns for items. Second, we expand on the selecting procedure of the Q-matrix with the highest posterior probability among all possible Q-matrices. Then, we formulate a recursive update rule for the proposed method using variational Bayes estimates and the selection procedure of an updated Q-matrix during the iteration. Finally, the issue of column permutation is discussed.

\subsubsection*{Likelihood and posterior probability of possible $q$-vector patterns of a Q-matrix}
Let $\mathbf{E}_j=(\bm{\epsilon}_{j1},\ldots,\bm{\epsilon}_{jh},\ldots,\bm{\epsilon}_{jH} )^{\top}$ be the $H \times K$ matrix of possible $q$-vector patterns of a given Q-matrix for item $j$, where $H=2^K-1$ represents the number of all possible $q$-vector patterns for this item. Let $\bm{\epsilon}_{jh}=(\epsilon_{jh1},\ldots,\epsilon_{jhk},\ldots,\epsilon_{jhK})^{\top}$ be the $h$-th row of $\mathbf{E}_j$. To treat the posterior probability of each possible $q$-vector pattern for item $j$, we convert each row of the matrix $\mathbf{E}_j$ to a decimal number, producing $H$ integers $\bm{\epsilon}_j^d=(\epsilon_{j1}^d,\ldots,\epsilon_{jh}^d,\ldots,\epsilon_{jH}^d )^{\top}$. Each entry of $\bm{\epsilon}_j^d$ coincides with the $h$-th row of the matrix $\mathbf{E}_j$. Furthermore, we assign the Bernoulli distribution $p(q_{jk}=1)=\gamma_{jk}$, whose prior is the Beta distribution $\gamma_{jk} \sim \mathrm{Beta}(1,1)$, to each element of a Q-matrix. Then, the conditional posterior for $\gamma_{jk}$ becomes $\mathrm{Beta}(1+q_{jk},2-q_{jk})$. By specifying $\gamma_{jk}$ for each element in a Q-matrix, we can define the $H \times K$ matrix $\bm{\Phi}_j=(\bm{\phi}_{j1},\ldots,\bm{\phi}_{jh},\ldots,\bm{\phi}_{jH})^{\top}$ for item $j$. Each entry of $\bm{\Phi}_j$ corresponds to the probability of taking the value of 1 at each entry of $\mathbf{E}_j$. Every $h$-th row vector of $\bm{\Phi}_j$ is set to be equal to $\bm{\gamma}_j=(\gamma_{j1},\ldots,\gamma_{jk},\ldots,\gamma_{jK})^{\top}$ such that the probability of each element of a Q-matrix is reflected in the probability of each possible $q$-vector pattern for item $j$. The prior distribution $p(\bm{\phi}_j^d)$ for each possible $q$-vector pattern for item $j$ is derived by transforming the matrix $\bm{\Phi}_j$ as
\begin{align*}
p(\bm{\phi}_j^d )&= \Bigl(p(\phi_{j1}^d ),p(\phi_{j2}^d ),\ldots,p(\phi_{jH}^d )\Bigl) \\
&= \left( \prod_{k=1}^K \phi_{j1k}^{\epsilon_{j1k}} (1-\phi_{j1k} )^{1-\epsilon_{j1k}}, \prod_{k=1}^K \phi_{j2k}^{\epsilon_{j2k}} (1-\phi_{j2k})^{1-\epsilon_{j2k}},\ldots, \prod_{k=1}^K \phi_{jHk}^{\epsilon_{jHk}} (1-\phi_{jHk})^{1-\epsilon_{jHk}} \right).
\end{align*}

Each entry of $p(\bm{\phi}_j^d )$ corresponds to the prior probability of each possible $q$-vector pattern for item $j$. The posterior probability $p(\phi_{jh}^d \vert \bm{x}_j )\propto p(\bm{x}_j \vert \phi_{jh}^d )p(\phi_{jh}^d )$ is estimated by the likelihood of $\phi_{jh}^d$ and the prior $p(\phi_{jh}^d)$. The likelihood $p(\bm{x}_j \vert \phi_{jh}^d)$ is given as 
\begin{align*}
p(\bm{x}_{j} \vert \phi_{jh}^d)= \prod_{i=1}^N \Bigl( (1-s_j)^{x_{ij}} s_j^{1-x_{ij}} \Bigl)^{\eta_{ih}} \Bigl(g_j^{x_{ij}}(1-g_j)^{1-x_{ij}} \Bigl)^{1-\eta_{ih}},
\end{align*}
where $\eta_{ih} = \prod_{k=1}^K \alpha_{ik}^{\epsilon_{hk}}$ is the ideal response. In the proposed method, the DINA model parameters $\{\mathbf{A}, s_j, g_j \}$ are estimated through VB inference and then imputed into the likelihood of $\phi_{jh}^d$ as fixed values. Thus, this quantity can be considered an item-wise marginal likelihood given the point VB estimates, which quantifies the model probability of observing $\bm{x}_{j}$ when the $h$-th row vector pattern of $\mathbf{E}_j$ is selected as a Q-matrix specification for item $j$.

\subsubsection*{Selecting the Q-matrix with the highest posterior probability from all possible Q-matrices}
In the previous section, we defined $p(\phi_{jh}^d \vert \bm{x}_j )$ as the posterior probability of each possible $q$-vector pattern for item $j$. Here, we expand on the selection procedure of an updated Q-matrix using $p(\phi_{jh}^d \vert \bm{x}_j )$. Consider the case with $K=2$ and $J=2$. Then, the number of all possible $q$-vector patterns $H=2^K-1$ becomes three, and the number of all possible Q-matrices $(2^K-1)^J$ becomes nine. Accordingly, three possible $q$-vector patterns and their posterior probabilities can be constructed for each item such that 
\begin{align*}
\text{For item 1}: \left(
\begin{array}{cc}
1 & 0 \\
0 & 1 \\
1 & 1
\end{array}
 \right) 
\begin{array}{c}
\rightarrow p(\phi_{11}^d \vert \bm{x}_1 )\\
\rightarrow p(\phi_{12}^d \vert \bm{x}_1 )\\
\rightarrow p(\phi_{13}^d \vert \bm{x}_1 )
\end{array}
\quad
\text{For item 2}: \left(
\begin{array}{ccc}
1 & 0 \\
0 & 1 \\
1 & 1
\end{array}
 \right)
 \begin{array}{c}
\rightarrow p(\phi_{21}^d \vert \bm{x}_2 )\\
\rightarrow p(\phi_{22}^d \vert \bm{x}_2 )\\
\rightarrow p(\phi_{23}^d \vert \bm{x}_2 )
\end{array}.
\end{align*}
With those posterior probabilities of possible $q$-vector patterns, we can compute the posterior probability of possible Q-matrices. For example, the posterior probability for one of the nine possible Q-matrices $\mathbf{Q}_{\mathrm{possible}} = \left(
\begin{array}{cc} 1 & 0 \\ 1 & 1 \end{array}\right) $ with $\bm{q}_1=(1, 0)$ and $\bm{q}_2=(1, 1)$ can be computed as $p(\mathbf{Q}_{\mathrm{possible}}\vert \mathbf{X}) = p(\phi_{11}^d \vert \bm{x}_1 )p(\phi_{23}^d \vert \bm{x}_2 )$. Hence, by selecting the possible $q$-vector pattern with the highest posterior probability for each item, the Q-matrix with those possible $q$-vector patterns readily becomes that with the highest posterior probability among all possible Q-matrices. Therefore, we do not need to evaluate posterior probabilities of all the possible Q-matrices to select the Q-matrix with the highest posterior probability. This Q-matrix can be easily found by simply selecting the possible $q$-vector pattern with the highest posterior probability for each item. Additionally, posterior probabilities of possible $q$-vector patterns are analytically evaluated, and their computation can be implemented quite efficiently by vectorizing the calculations. Note that the posterior probability of a possible Q-matrix $p(\mathbf{Q}_{\mathrm{possible}} \vert \mathbf{X})$ can be regarded as the posterior \emph{model} probability given the data $\mathbf{X}$ and the point VB estimates because a Q-matrix is regarded as a structural model component in the proposed method.

\subsubsection*{Formulating a recursive update rule of a Q-matrix}
Let $\mathbf{Q}^t$ be the Q-matrix at the $t$-th iteration, and let us consider the VB estimates of slip $\bm{s}^t$, guessing $\bm{g}^t$, and attribute mastery profiles $\mathbf{A}^t$ given $\mathbf{Q}^t$. These estimates are the expected a posteriori (EAP) estimates from the variational posterior distributions $q^t(\bm{s}_j)$, $q^t (\bm{g}_j)$ and the maximum a posteriori (MAP) estimates from $q^t (\bm{z}_{\tau_s})$ for all $\tau_s$. The attribute mastery profiles $\mathbf{A}^t$ are computed by selecting the attribute mastery pattern that corresponds to the position containing the largest value in the vector $\bm{z}_{\tau_s}$ for each of $S$ individuals. As there is a one-to-one correspondence between the VB estimates calculated from the variational posteriors and the ELBO at the $t$-th iteration, we can regard these estimates as the quantity reflecting the parameter values that best fit the data at the current $\mathbf{Q}^t$. By substituting these VB estimates into the likelihood of $\bm{\phi}_{j}^d$, the information of $\mathbf{Q}^t$ is carried over to the selection procedure for choosing the Q-matrix with the highest posterior probability. Hence, the updated Q-matrix $\mathbf{Q}^{t+1}$ is expected to produce a larger ELBO than the one at $\mathbf{Q}^t$. This idea is similar to the case in which the gradient of an objective function guides the direction that produces the largest change in the value of that function. Although this selection procedure runs deterministically, stochastic variability emerges from stochastically choosing \textcolor{black}{the mini-batch samples} $\mathbf{X}_S$ at every iteration. This stochasticity facilitates an effective search for the optimal elements of the Q-matrix.

However, owing to this variability, a sequence of Q-matrices never converges to a single Q-matrix. For this issue, we employ iterate averaging to determine the final estimate. That is, we use the average of updated Q-matrices in the iteration as the final estimate of a Q-matrix. This idea originates from \textcite{polyak_new_1990} and \textcite{ruppert_efficient_1988}, and it is referred to as Polyak-Ruppert averaging \parencite{neu_iterate_2018,polyak_acceleration_1992}. \textcolor{black}{Although Polyak-Ruppert averaging was introduced for accelerating and stabilizing stochastic approximation algorithms, we found that the averaging procedure works well in the proposed method.} Since an element of a Q-matrix is binary, we round the average of the updated Q-matrices, resulting in taking the mode in a sequence of the updated Q-matrices. We observed from the simulation study that the final estimate becomes an optimal solution when the sequence of Q-matrices is from the neighborhood of the true Q-matrix. Note that despite $(2^K-1)^J$ Q-matrices being considered for each update, the parameter space of an iterate-averaged Q-matrix includes Q-matrices with items that do not measure any of the attributes. Accordingly, possible Q-matrices for the final estimate amount to $2^{JK}$. If the iterate-averaged Q-matrix possesses such an item, it is interpreted as a residual that cannot be explained by the given attributes. In practice, we discard the first $D$ iterates and use the rest of the updated Q-matrices for iterate averaging. 

Moreover, we run this iteration algorithm for the predefined number of times $R$; calculate the mean of the $T-D$ ELBOs for each run, whose values correspond to each of the updated Q-matrices; and select the estimated Q-matrix from the run with the largest mean ELBOs as the final estimate. The entire estimation algorithm that constitutes the proposed method is summarized in the pseudocode (Table \ref{table:pseudocode}).

\begin{table}
\caption{Pseudocode for the proposed method}
\begin{algorithm}[H]
\caption{Estimation procedure of the proposed algorithm}
\begin{algorithmic}
\STATE Given $R$, $T$, $D$, $S$, $\mathbf{Q}^0$
\FOR{ $r = 1$ to $R$}
\FOR{ $t = 1$ to $T$}
\STATE Sample $S$ indices $\tau_s \sim \mathrm{DiscreteUniform}(1, N)$ to construct $\mathbf{X}^t_S$
\STATE Optimize $L(q^t, \mathbf{X}^t_S, \mathbf{Q}^{t-1})$
\STATE Obtain the estimates of slip $\bm{s}^t$, guessing $\bm{g}^t$, and attribute mastery profile $\mathbf{A}^t_S$
\STATE Calculate $p^t(\phi^d_{jh} \vert \bm{x}_j^t)$ given $s^t_j$, $g^t_j$, and $\mathbf{A}^t_S$ for all the combinations of $j$ and $h$
\STATE Set $\mathbf{Q}^t \leftarrow \argmax_{\mathbf{Q}} p(\mathbf{Q} \vert \mathbf{X}^t_S, \bm{s}^t, \bm{g}^t, \mathbf{A}^t_S)$
\ENDFOR
\STATE Discard the first $D$ iterates
\STATE Compute $\hat{\mathbf{Q}}_r$ by averaging $T-D$ Q-matrices with rounding
\ENDFOR
\STATE Select $\hat{\mathbf{Q}}_r$ with the largest mean of the $T-D$ ELBOs as the final estimate $\hat{\mathbf{Q}}$
\RETURN $\hat{\mathbf{Q}}$
\end{algorithmic}
\end{algorithm}
\label{table:pseudocode}
\end{table}

\subsubsection*{Column permutation}
When estimated from data, a Q-matrix can be identified up to a column permutation. This means that data itself cannot distinguish the permutation of the columns of a Q-matrix because $K!$ combinations of Q-matrices produce equal likelihood \parencite{liu_theory_2013}. Hence, label switching has been a major problem in MCMC-based Bayesian Q-matrix estimation when summarizing the estimate of $\mathbf{Q}$. However, the proposed method does not encounter the label-switching problem for two reasons. First, although this method implements multiple runs of the iterative algorithm similar to chains of the MCMC-based Bayesian Q-matrix estimation, we do not combine the updated Q-matrices from these runs for estimating a Q-matrix. Instead, we compute its estimate using the updated Q-matrices only from one run that produced the largest mean of ELBOs. Thus, the label switching induced by different attribute labels of initial Q-matrices \emph{between} multiple runs does not occur. Second, as the proposed method is essentially deterministic, the label switching \emph{within} a run also does not emerge. In our method, given a Q-matrix associated with one specific attribute label among $K!$ attribute-label patterns, the estimates of the DINA model parameters are obtained through variational inference. Then, to compute the posterior probabilities of possible $q$-vector patterns for items, we imputed the estimates of attribute mastery profiles and item parameters into the likelihood of possible $q$-vector patterns. The imputed mastery profiles hold the same attribute label as the Q-matrix given in the previous step for estimating the DINA model parameters. Thus, an attribute label in the likelihood of possible $q$-vector patterns is the same as the Q-matrix in the previous step. Theoretically, a set of the likelihood's values for possible $q$-vector patterns can be equivalent across $K!$ patterns of attribute labels. Nonetheless, this occurs only when an attribute label on both attribute mastery profiles and possible $q$-vector patterns (or a Q-matrix) changes correspondingly. The proposed method calculates the likelihood of possible $q$-vector patterns by imputing the estimated mastery profiles as fixed values and deterministically selects the Q-matrix with the highest posterior probability under the attribute label of the imputed mastery profiles. Hence, an attribute label on both attribute mastery profiles and possible $q$-vector patterns never changes correspondingly during the iteration. It should be noted that different \textcolor{black}{mini-batch samples} during the iteration do not cause the label switching. This is because the label on attribute mastery profiles in these data is determined by a Q-matrix, and not by the data itself. These data contribute only to the likelihood's values for possible $q$-vector patterns under the attribute label of the imputed mastery profiles.

Therefore, our method needs no relabeling of the columns for the $T-D$ updated Q-matrices after performing multiple runs of the iterative algorithm. We can simply average the $T-D$ updated Q-matrices without taking care of label switching. This is a desirable property, especially when determining the EAP estimate of a Q-matrix with a large number of attributes, because the relabeling algorithm for calculating the EAP estimate in Bayesian Q-matrix estimation requires the evaluation of the distance between the reference matrix and $K!$ combinations of the Q-matrix being relabeled \parencite{chung_gibbs_2019}, which leads to a computational burden. If the reference matrix is provided in the form of a provisional Q-matrix by domain experts, it suffices to relabel only an estimated Q-matrix to reflect the prior information of each attribute label onto the estimate.

\section{Simulation Study}
\subsection{Simulation Procedure}
To verify the effectiveness of the proposed algorithm, we conducted a simulation study under small- and large-scale conditions. The simulation settings were as follows. For the small-scale conditions, we considered the number of attributes $K=3$ and $4$; the sample size of $250$, $500$, and $1000$; and the number of items $J=10$ and $20$. For the large-scale conditions, we considered the number of attributes $K=7$ and $8$; the sample size of $2000$, $4000$, and $8000$; and the number of items $J=40$ and $80$. The detailed specifications of the true Q-matrices for this simulation study are provided in Supplementary Material A. The correlation coefficient among attributes was set to either $\rho=0$ or $0.25$. The slip and guessing parameters were set to $0.2$ in all the conditions. A total of 100 datasets were generated for each combination of the Q-matrix, sample size, and correlation conditions.

To generate correlated attributes, we followed the same procedure used in \textcite{chung_gibbs_2019}. As the correlation between each pair of attributes is assumed to be identical for all pairs in this simulation, the $K \times K$ correlation matrix $\bm{\Sigma}$ is given as
\begin{align*}
\bm{\Sigma} = \left(
\begin{array}{ccc}
1 & \cdots& \rho \\
\vdots & \ddots & \vdots \\
\rho & \cdots& 1
\end{array}
 \right).
\end{align*}
Here, the common correlation coefficient $\rho$ is assigned to all the off-diagonal entries. Since $\bm{\Sigma}$ is a real symmetric positive-definite matrix, we apply the Cholesky decomposition $\bm{\Sigma}=\mathbf{V}^{\top}\mathbf{V}$, where $\mathbf{V}$ is an upper triangular matrix. Furthermore, an $N \times K$ matrix $\mathbf{T}$ whose entries are generated from the standard normal distribution $N(0,1)$ is created. Then, this $\mathbf{T}$ is transformed as $\bm{\nu}=\mathbf{TV}$ in order for $\bm{\nu}$ to have the same correlation structure as $\bm{\Sigma}$. Subsequently, let $\bm{
\Lambda}$ be the $N \times K$ matrix representing the underlying probability of $\alpha_{ik}$, and let us set $\phi(\bm{\nu})=\bm{\Lambda}$, where $\phi(\cdot)$ is the cumulative density function of the standard normal distribution. To generate $\alpha_{ik}$, we apply the following criteria \parencite{chiu_cluster_2009, liu_data-driven_2012}:
\begin{align}
\alpha_{ik} = \begin{cases}
1 \;\text{if}\; \lambda_{ik} \geq \phi^{-1}\left(\frac{k}{K+1}\right) & \\
 0 \;\text{otherwise} &
\end{cases}.
\end{align}
Eq. (1) assumes that each attribute has different difficulty levels for its mastery.

Regarding item responses, we employed the inverse transform sampling for Bernoulli random variables \parencite{ross_simulation_2013}. First, the ideal response for each examinee $\eta_{ij}$ is calculated from the generated attributes and Q-matrix. Second, we calculate the $N \times J$ probability matrix $\mathbf{P}$ of correct responses for all the combinations of examinees and items by using the predefined slip and guessing parameters. Third, we generate another $N \times J$ probability matrix $\mathbf{C}$ with each entry sampled from $\mathrm{Uniform}(0,1)$. Then, the elements of the probability matrices $\mathbf{P}$ and $\mathbf{C}$ are compared, and the item response $x_{ij}$ is given as
\begin{align*}
x_{ij} = \begin{cases}
1 \;\text{if}\; p_{ij} \geq c_{ij} & \\
0 \;\text{otherwise} &
\end{cases}.
\end{align*}

To assess the performance of the proposed method, we calculated the element- and matrix-wise mean recovery rates to determine the accuracy for estimating a Q-matrix. Specifically, let $\hat{\mathbf{Q}}^w=(\hat{q}_{jk}^w)_{J \times K}\;(w=1,\ldots,W)$ and $\mathbf{Q}^{\mathrm{true}}=(q_{jk}^{\mathrm{true}})_{J \times K}$ be the estimated Q-matrix from the $w$-th dataset and the true Q-matrix, respectively. The element-wise mean recovery rate (eMRR) is defined as
\begin{align*}
\mathrm{eMRR} = \frac{1}{W}\sum_{w=1}^W \left(1 - \frac{\sum_{j=1}^J \sum_{k=1}^K \vert \hat{q}_{jk}^w - q_{jk}^{\mathrm{true}} \vert }{JK} \right),
\end{align*}
where $\vert \cdot \vert $ indicates the absolute value.
The matrix-wise mean recovery rate (mMRR) is defined as 
\begin{align*}
\mathrm{mMRR} = \frac{1}{W}\sum_{w=1}^W I\left(\hat{\mathbf{Q}}^w = \mathbf{Q}^{\mathrm{true}}\right), 
\end{align*}
where $I(\cdot)$ denotes the indicator function that takes the value of 1 when the given condition is satisfied.

Finally, four hyperparameters must be set when running the proposed method: \textcolor{black}{the mini-batch} sample size $S$ for the stochastic optimization, the number of times to run the iteration algorithm $R$, the number of iterations in a single run $T$, and the number of discarded iterations for iterate averaging $D$. For the simulation study, \textcolor{black}{the mini-batch} sample size was set to $S=200$ for $N=250$ and $S=300$ for $N=500$, $1000$, $2000$, $4000$, and $8000$. Additionally, we set $R=10$, $T=550$, and $D=50$ for the small-scale conditions and set $R=10$, $T=650$, and $D=150$ for the large-scale conditions. Regarding the setting of the VB estimation for the DINA model parameters, we stopped the iteration when the maximum change in successive ELBO's value became less than $10^{-5}$, or when the number of iterations reached 1,000. This specification for the VB estimation is the same as that employed in \textcite{yamaguchi_variational_2020}. The initial Q-matrices for each simulation dataset were randomly generated. The program was written and implemented with parallel computing in the Julia programming language \parencite[version 1.54;][]{bezanson_julia_2017}, wherein we parallelized the computation using the \texttt{Distributed} standard library in Julia. The Julia code for this article is available on the Open Science Framework \url{https://osf.io/jev9q/?view_only=e1b1f047c89f46cba9a3b61194404d8e}

\subsection{Results}

The element- and matrix-wise mean recovery rates for each condition are presented in Table \ref{table:results_Q_sim}. Regarding the small-scale conditions with $K=3$ and $4$, in general, the increase in the number of items positively contributed to an accurate estimation, and that in correlation coefficient among attributes slightly deteriorated the estimation. The element-wise mean recovery rates were beyond 93\% and satisfactory across almost all the conditions, except for the small-scale ones with $K=4$, $J=10$, and $N=250$ under $\rho=0$ and with $K=4$, $J=10$, and $N=250$ and $500$ under $\rho=0.25$. In these conditions, the element-wise rates were 85--90\%. This insufficient performance under such small-scale conditions would be attributed to the insufficient number of items and samples. However, when the sample size increased to $1000$, the element-wise rates became over 93\%. This indicates that increasing sample size can compensate for the lack of items even in a small Q-matrix for four attributes. The matrix-wise mean recovery rates under the small-scale conditions were also high especially under the conditions with $K=3$. In particular, as the sample size increased, the proposed method recovered the true Q-matrices in almost all the replications under $K=3$. When $K=4$, the full recovery of the true Q-matrices was more difficult than that of $K=3$, although the increase in the sample size positively contributed to better performance in the full Q-matrix recovery.

Regarding the large-scale conditions with $K=7$ and $8$, the increase in the number of items had a positive effect \textcolor{black}{for $K=8$ and a negative one for $K=7$} on an accurate estimation. Interestingly, the increase in the correlation coefficient among attributes also had a positive impact on the estimation. This would be because the correlation among attributes with high dimensionality is positively associated with estimating a $q$-vector measuring multiple attributes. In addition, the element-wise mean recovery rates were beyond 94\% and satisfactory across all the conditions. The full recovery of the true Q-matrices in these conditions was more difficult than that in the small-scale condition because the number of elements in a Q-matrix is much larger. Nonetheless, most of the conditions achieved more than approximately 97\% in terms of the element-wise mean recovery rate.

\begin{table}[!htbp]
\color{black}
\centering
\caption{Mean recovery rate under each condition in the simulation study}
 \resizebox{\linewidth}{!}{
\begin{tabular}{CCCRRRCCCRR}
 \toprule
 \multicolumn{11}{c}{$\rho=0$} \\
 \midrule
 K& J& N& \multicolumn{1}{c}{matrix-wise} & \multicolumn{1}{c}{element-wise} & & K& J& N& \multicolumn{1}{c}{matrix-wise} & \multicolumn{1}{c}{element-wise} \\
\cmidrule{1-5}\cmidrule{7-11} 3 & 10& 250 & 41/100 & 97.00\% & & 7 & 40& 2000& 2/100 & 97.71\% \\
& & 500 & 80/100 & 99.17\% & & & & 4000& 13/100 & 98.93\% \\
& & 1000& 91/100 & 99.67\% & & & & 8000& 27/100 & 98.89\% \\
& 20& 250 & 57/100 & 98.97\% & & & 80& 2000& 4/100 & 95.83\% \\
& & 500 & 93/100 & 99.87\% & & & & 4000& 16/100 & 96.31\% \\
& & 1000& 100/100 & 100.00\% & & & & 8000& 19/100 & 96.25\% \\
\cmidrule{1-5}\cmidrule{7-11}4 & 10& 250 & 5/100 & 87.18\% & & 8 & 40& 2000& 0/100 & 94.88\% \\
& & 500 & 15/100 & 93.38\% & & & & 4000& 1/100 & 95.14\% \\
& & 1000& 35/100 & 97.30\% & & & & 8000& 3/100 & 95.83\% \\
& 20& 250 & 4/100 & 95.84\% & & & 80& 2000& 3/100 & 98.48\% \\
& & 500 & 46/100 & 98.82\% & & & & 4000& 19/100 & 98.95\% \\
& & 1000& 76/100 & 99.62\% & & & & 8000& 31/100 & 99.21\% \\
 \midrule
 & & & & & & & & & &\\
 \toprule
 \multicolumn{11}{c}{$\rho=0.25$} \\
 \midrule
 K& J& N& \multicolumn{1}{c}{matrix-wise} & \multicolumn{1}{c}{element-wise} & & K& J& N& \multicolumn{1}{c}{matrix-wise} & \multicolumn{1}{c}{element-wise} \\
\cmidrule{1-5}\cmidrule{7-11} 3 & 10& 250 & 25/100 & 94.33\% & & 7 & 40& 2000& 5/100 & 98.54\% \\
& & 500 & 52/100 & 97.43\% & & & & 4000& 16/100 & 99.13\% \\
& & 1000& 82/100 & 99.30\% & & & & 8000& 35/100 & 99.48\% \\
& 20& 250 & 39/100 & 98.22\% & & & 80& 2000& 3/100 & 97.33\% \\
& & 500 & 77/100 & 99.50\% & & & & 4000& 28/100 & 98.48\% \\
& & 1000& 93/100 & 99.88\% & & & & 8000& 54/100 & 98.79\% \\
\cmidrule{1-5}\cmidrule{7-11}4 & 10& 250 & 0/100 & 85.15\% & & 8 & 40& 2000& 0/100 & 96.33\% \\
& & 500 & 3/100 & 90.12\% & & & & 4000& 1/100 & 96.77\% \\
& & 1000& 18/100 & 93.87\% & & & & 8000& 5/100 & 97.84\% \\
& 20& 250 & 2/100 & 93.64\% & & & 80& 2000& 1/100 & 97.48\% \\
& & 500 & 16/100 & 97.10\% & & & & 4000& 1/100 & 98.53\% \\
& & 1000& 43/100 & 98.68\% & & & & 8000& 19/100 & 99.14\% \\
 \bottomrule
 \end{tabular}%
 }
\label{table:results_Q_sim}%
\end{table}%

\begin{figure}[!htbp]
\color{black}
 \begin{center}
 %\begin{tabular}{c}
 \begin{minipage}{0.8\linewidth}
 \centering
 \includegraphics[width=\linewidth]{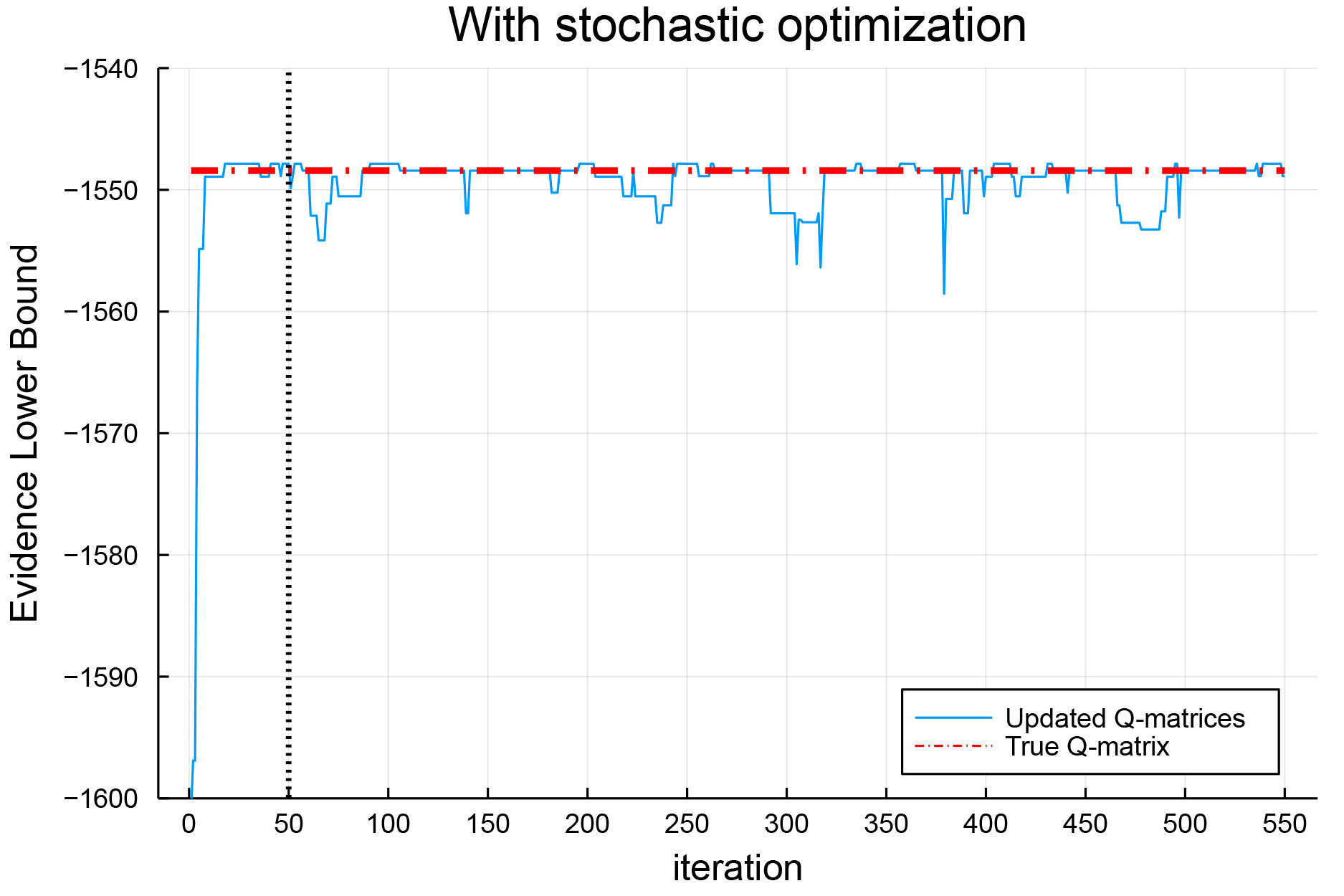}
 \end{minipage} 
 \begin{minipage}{0.8\linewidth}
 \centering
 \includegraphics[width=\linewidth]{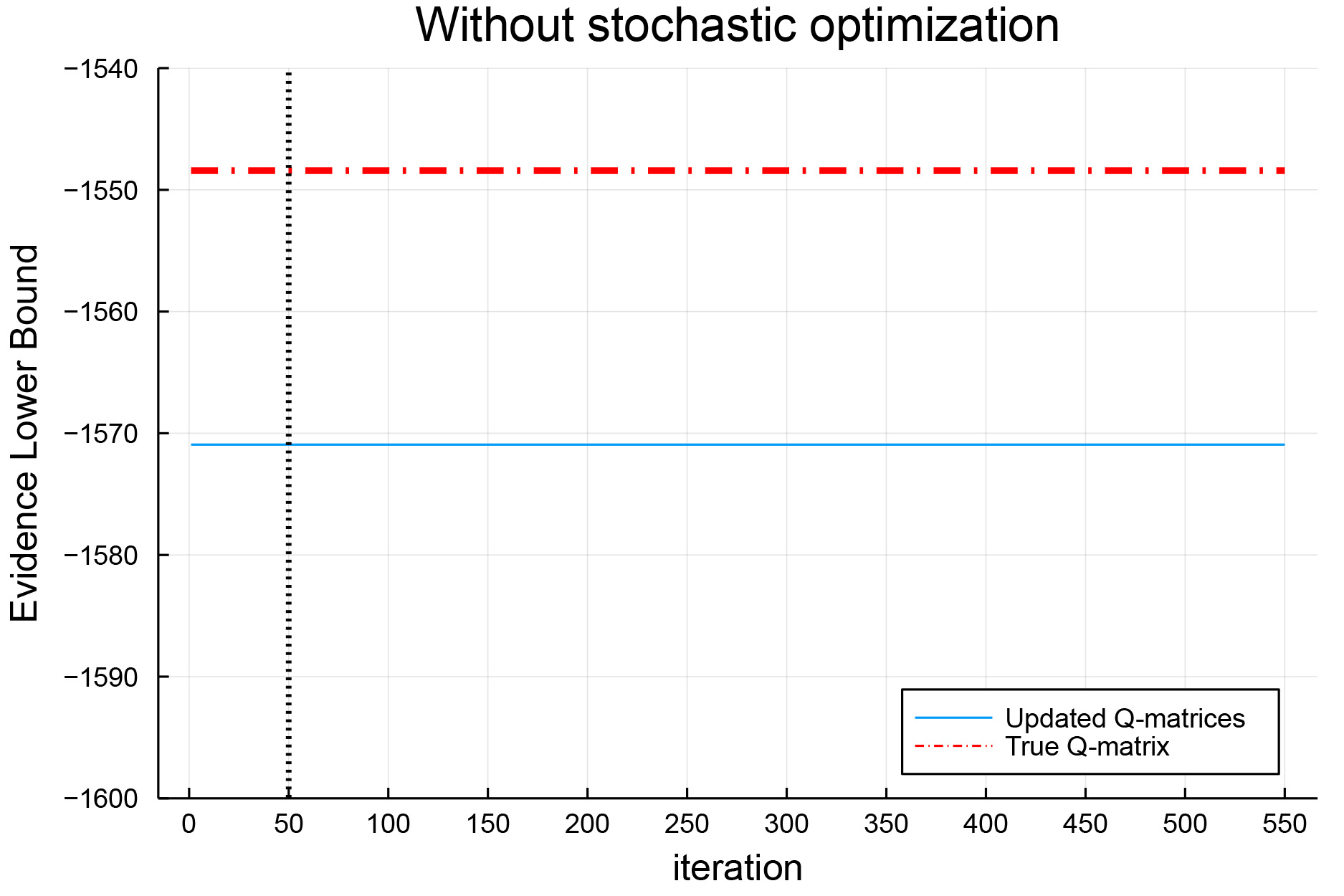}
 \end{minipage}
 %\end{tabular}
\caption{The transition of the ELBOs in one of the simulation datasets when $K=3$ and $J=10$, the sample size is $250$, and correlation coefficient is $0$. The $x$-axis represents the number of iterations, and the $y$-axis represents the ELBO with all data points given a Q-matrix. The red line indicates the value of the ELBO given the true Q-matrix. The blue line indicates the transition of the ELBOs given the updated Q-matrices. While the proposed algorithm with stochastic optimization recovered 100\% of the true Q-matrix elements, the algorithm without optimization recovered 73.33\% of its elements. Upper panel: proposed algorithm equipped with stochastic optimization; lower panel: the same algorithm without stochastic optimization.}
\label{fig:transition_ELBO_N500}
 \end{center}
\end{figure}

\begin{figure}[!htbp]
\color{black}
 \begin{center}
 %\begin{tabular}{c}
 \begin{minipage}{0.8\linewidth}
 \centering
 \includegraphics[width=\linewidth]{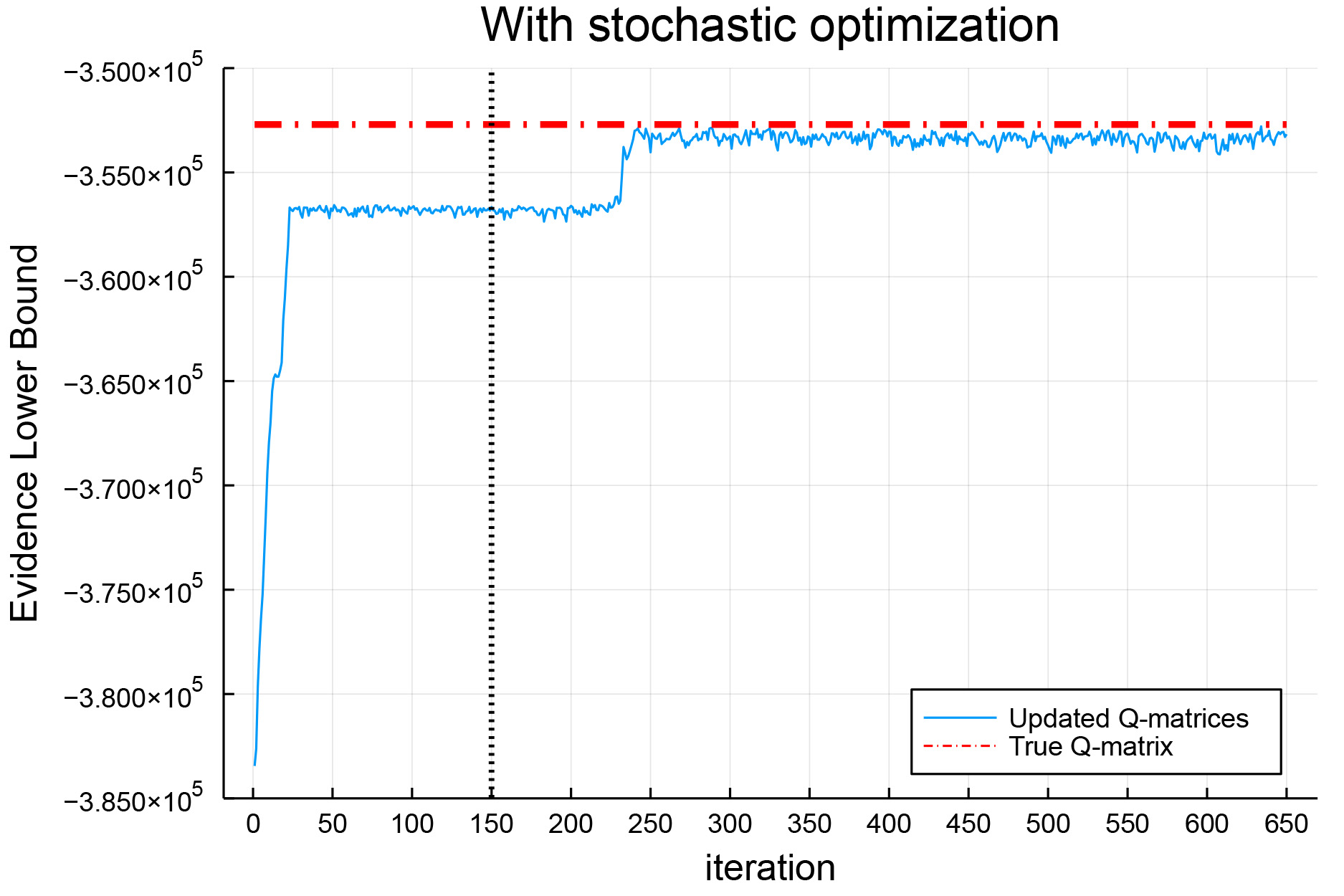}
 \end{minipage} 
 \begin{minipage}{0.8\linewidth}
 \centering
 \includegraphics[width=\linewidth]{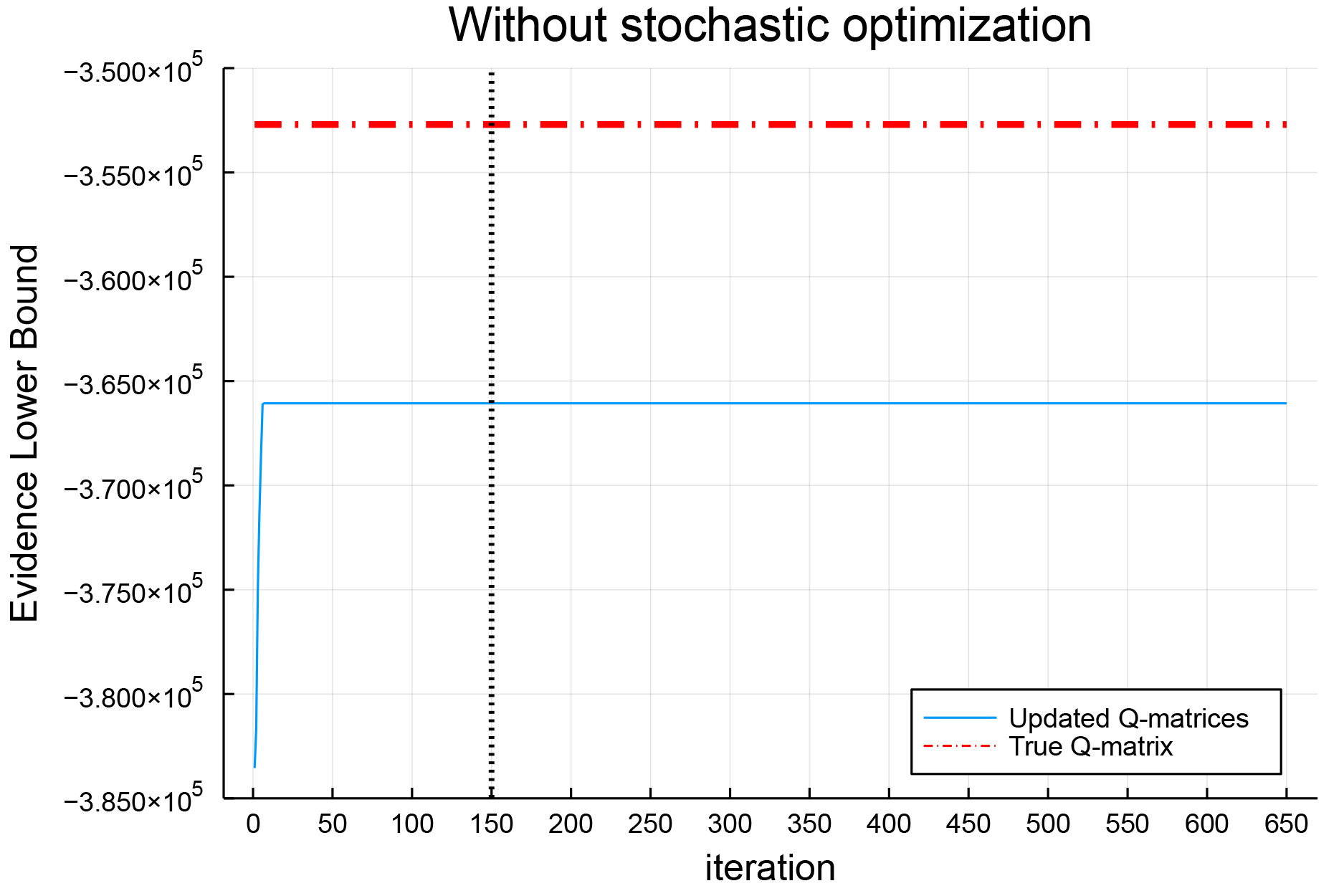}
 \end{minipage}
 %\end{tabular}
\caption{The transition of the ELBOs in the first simulation dataset when $K=8$ and $J=80$, the sample size is $8000$, and correlation coefficient is $0$. While the proposed algorithm with stochastic optimization recovered 100\% of the true Q-matrix elements, the algorithm without stochastic optimization recovered 74.38\% of its elements. Notations are the same as in Figure 2.}
\label{fig:transition_ELBO_N2000}
 \end{center}
\end{figure}

To illustrate how stochastic variability derived from stochastic optimization facilitates an effective search for the optimal Q-matrix, Figures \ref{fig:transition_ELBO_N500} and \ref{fig:transition_ELBO_N2000} represent the transition of the ELBOs during the iteration in one of the datasets from two of the simulation conditions of $K=3$ and $K=8$. The same simulation dataset was analyzed with both the proposed algorithm that is equipped with stochastic optimization and the same algorithm without stochastic optimization. The values of the ELBOs were recalculated given all data points after a selected sequence of the updated Q-matrices had been obtained from the run with the largest mean of the ELBOs. It is evident from the figures that shuffling \textcolor{black}{the mini-batch samples from} all data points at every iteration produces stochastic variability bringing the updated Q-matrices to an optimal area of Q-matrices. Hence, stochastic optimization not only reduces computational complexity but also contributes to the effectiveness of optimization.

\section{Empirical Study}
To demonstrate the utility of the proposed method, three empirical studies were conducted using the fraction subtraction, Examination for the Certificate of Proficiency in English (ECPE), and Trends in International Mathematics and Science Study (TIMSS) 2003 mathematics datasets. Each of the empirical studies presumes three types of real field settings: the fraction subtraction dataset is characterized as a setting with a moderate number of attributes and small sample size, the ECPE dataset is a setting with a small number of attributes and large sample size, and the TIMSS 2003 mathematics dataset is a setting with a considerably large number of attributes and large sample size. From the viewpoint of five relative model fit indices (i.e., Akaike information criterion (AIC), Bayesian information criterion (BIC), widely applicable information criterion \parencite[WAIC;][]{watanabe_asymptotic_2010}, widely applicable Bayesian information criterion \parencite[WBIC;][]{watanabe_widely_2013}, and Negative ELBO) and computation time, we compared the Q-matrix estimated by the proposed method with other Q-matrices estimated using two existing methods: the Gibbs sampler \parencite{chung_gibbs_2019} and EM-based algorithm with Lasso regularization \parencite{chen_statistical_2015}. \textcolor{black}{Specifically, after estimating a Q-matrix with these three methods, we treated each of the estimated Q-matrices as a known component in the same way that practitioners usually do in a DCM analysis. (For example, they often consider an expert-defined Q-matrix to be a known component in structural and item parameter estimation for their chosen model.) Subsequently, we applied a certain type of estimation method to the data to obtain estimates of the DINA model parameters, which are necessary to compute model fit measures, given the corresponding estimated Q-matrices. Then, we computed the values of relative model fit indices for each estimated Q-matrix. In the case of WAIC and WBIC, we used a Bayesian MCMC estimation method (a Gibbs sampler for the DINA model parameter estimation, not for Q-matrix estimation) for the analysis. Similarly, we used an EM-algorithm in the case of AIC and BIC and a Bayesian variational estimation method in the case of ELBO.} In addition, as the methods do not enforce identification constraints in the estimation procedure, an estimated Q-matrix can hold an unidentified structure for structural and item parameters. If this occurs, consistent estimates for these parameters cannot be obtained with the unidentified Q-matrix. \parencite{fang_identifiability_2019, gu_partial_2020, gu_sufficient_2019, chen_statistical_2015, xu_identifiability_2017, xu_identifying_2018, xu_identifiability_2016}. Thus, we analyzed the identifiability of the estimated Q-matrices by investigating whether the estimated Q-matrices satisfy the conditions of strict and partial identifiability for the DINA model \parencite{gu_partial_2020}.

\subsection{Data Descriptions}
\subsubsection{Fraction Subtraction Dataset}
This dataset contains binary responses from 536 middle school students to 17 fraction subtraction items. Various versions of this dataset have been analyzed in the DCM literature \parencite{de_la_torre_higher-order_2004,delatorre_generalized_2011,tatsuoka_data_2002}. We adopted the same version used in \textcite{chen_statistical_2015} and \textcite{xu_identifying_2018}, which is available on the \texttt{CDM} R package \parencite{robitzsch_cdm_2020}. The specification of attributes for the fraction subtraction dataset is presented by \textcite{de_la_torre_higher-order_2004} as follows: (A1) Convert a whole number to a fraction, (A2) Separate a whole number from a fraction, (A3) Simplify before subtracting, (A4) Find a common denominator, (A5) Borrow from the whole number part, (A6) Column borrow to subtract the second numerator from the first, (A7) Subtract numerators, and (A8) Reduce answers to the simplest form. Previous studies using this dataset to estimate a Q-matrix found that assigning eight attributes is too detailed for this assessment and adopted a smaller number of attributes \parencite[e.g., $K=3,4,5$;][]{chen_statistical_2015,xu_identifying_2018}. Hence, we performed an exploratory estimation of the Q-matrix with $K=4$. Then, we compared our estimate with the Q-matrices estimated from the Gibbs sampler \parencite{chung_gibbs_2019} and EM-based algorithm \parencite{chen_statistical_2015}. The initial Q-matrices for the proposed method and Gibbs sampler were randomly generated. For the EM-based algorithm, we randomly generated the initial values for the intercepts and coefficients of attribute main- and interaction-effect terms from the uniform distribution $\mathrm{Uniform}(0,1)$. These initial values satisfied the monotonicity constraint on correct-response probabilities for different mastery profiles of required attributes. We also set the initial values of all the structural parameters to $\frac{1}{2^K}$ for the EM-based algorithm.

\subsubsection{ECPE Dataset}
The Examination for the Certificate of Proficiency in English (ECPE) is a standardized assessment for gauging advanced English skills of nonnative English students \parencite{templin_obtaining_2013}. This dataset was formerly analyzed with DCMs by \textcite{templin_hierarchical_2014} and \textcite{feng_parameter_2014}. It can be obtained from the \texttt{CDM} R package \parencite{robitzsch_cdm_2020}. Further, it consists of responses from 2922 examinees to 28 multiple-choice items extracted from the grammar section of the ECPE. These items measure the three attributes of morphosyntactic, cohesive, and lexical grammar. We performed an exploratory estimation of the Q-matrix with $K=3$ and compared our result with the estimates from the Gibbs sampler \parencite{chung_gibbs_2019} and EM-based algorithm \parencite{chen_statistical_2015}. 
In addition, because the EM-based algorithm with randomly generated initial values produced an estimated Q-matrix with some zero $q$-vectors, we set the initial values of this algorithm to the ML estimates of the DINA model parameters given the expert-defined Q-matrix in \textcite{templin_hierarchical_2014}. Accordingly, we set the expert-defined Q-matrix as the initial Q-matrices for the proposed method and Gibbs sampler and included it to prior distributions for both methods.

\subsubsection{TIMSS 2003 Mathematics Dataset}
Trends in International Mathematics and Science Study (TIMSS) is an international assessment of the mathematics and science achievement of fourth- and eighth-grade students. Its data have been garnered in a four-year cycle and utilized for designing effectual educational policy around the world. The subset of the U.S. eighth-grade mathematics data collected in 2003 was analyzed in DCM studies \parencite{skaggs_grain_2016,su_hierarchical_2013}, and we conducted an analysis using the dataset found in \textcite{su_hierarchical_2013}. A total of 757 examinees answering 23 items associated with 13 attributes comprise this dataset. The item response data are available on the \texttt{CDM} R package \parencite{robitzsch_cdm_2020}. We implemented the Q-matrix estimation by the proposed method and compared its performance with the Q-matrix specified by the domain experts in \textcite{su_hierarchical_2013}. This Q-matrix was also used as the initial value for the proposed method so that we do not need the computationally intensive relabeling procedure for the estimated Q-matrix. It should be noted that, although we performed Q-matrix estimation with the EM-based algorithm, this algorithm produced an estimated Q-matrix with the same $q$-vector for all the items because of the high-dimensional attribute main- and interaction-effect terms. Thus, we did not include the EM-based algorithm in this real data analysis.

\subsection{Settings for the Estimation}
For the analysis of the fraction subtraction dataset, \textcolor{black}{we set the four hyperparameters to $S=300$, $R=10$, $T=550$, and $D=50$. For the ECPE, we set them to $S=1200$, $R=10$, $T=550$, and $D=50$, and for the TIMSS 2003 mathematics, we assigned $S=400$, $R=10$, $T=550$, and $D=50$. Hyperparameters were set differently across different datasets to avoid such Q-matrix estimates that possess irregular items by which no attribute is measured, contain irregular attributes with which no measurement item is associated, or exhibit a substantially poor model fit.} The same setting specified in the simulation study was also applied for the VB estimation of the DINA model parameters. The analyses were performed on a desktop computer with 12-core AMD Ryzen 9 3900X and 64 GB memory. The Gibbs sampler \parencite{chung_gibbs_2019} and EM-based algorithm \parencite{chen_statistical_2015} used in the fraction subtraction and ECPE datasets were originally written in the R program \parencite{r_development_core_team_r_2020}. Accordingly, we re-implemented them as Julia code for better comparability. Following the original specification by \textcite{chung_gibbs_2019}, we ran three chains of 100,000 iterations with 50,000 burn-in parallelly by using three cores for the Gibbs sampler. For the EM-based algorithm, we ran this algorithm with 10 different tuning parameter values, parallelized these 10 implementations with 10 cores, and selected the Q-matrix estimate that yields the best-fitted value of the objective function as a final estimate. For the proposed method, we parallelized the computation of 10 runs ($R=10$) with 10 cores. It should be noted that, although performing 10 chains of the Gibbs sampler with 10 cores is theoretically possible, this estimation would be infeasible because implementing these 10 chains with 100,000 iterations requires a substantial amount of memory usage. Additionally, from the results of the simulation and empirical studies in \textcite{chung_gibbs_2019}, its original specification would suffice for an accurate estimation. Thus, we set three chains for the Gibbs sampler. The parallel computing was implemented by the \texttt{Distributed} standard library in Julia.

\subsection{Results}
\subsubsection{Fraction Subtraction Dataset}
The Q-matrices used for the comparison in this dataset are provided in Table B1 of Supplementary Material B. We compared the Q-matrix obtained from the proposed method with the Q-matrices estimated by the Gibbs sampler and EM-based algorithm. Table \ref{table:results_Fraction} shows the results. The proposed method surpasses the other two Q-matrices in terms of the five relative model fit indices. Furthermore, its computation time was approximately 52 times faster than that of the Gibbs sampler and comparable with the EM-based algorithm. In addition, the estimated Q-matrix from our method satisfies the strict identifiability for the DINA model, whereas other estimated Q-matrices do not satisfy it, but hold the $\bm{p}$-partial identifiability. This implies that these $\bm{p}$-partially identifiable Q-matrices cannot ensure consistent estimates of structural parameters for some attribute mastery patterns inseparable by the Q-matrices. The estimated Q-matrix from our method matched 94.12\% elements of the Q-matrix from the Gibbs sampler and matched 95.59\% elements of the Q-matrix from the EM algorithm.

\begin{table}[!htbp]
\color{black}
 \centering
 \caption{Fraction subtraction dataset: Relative model fit indices and computation time}
 \resizebox{\linewidth}{!}{
 \begin{tabular}{r C C C}
 \toprule
 &\multicolumn{1}{c}{Proposed method} & \begin{tabular}{c} Gibbs sampler \\\parencite{chung_gibbs_2019} \end{tabular} & \multicolumn{1}{c}{\begin{tabular}{c} EM-based algorithm \\ \parencite{chen_statistical_2015} \end{tabular} } \\
 \midrule
 AIC & 7460.52 & 7520.70 & 7489.86\\
 BIC & 7670.44 & 7730.62 & 7699.78 \\
 WAIC & 6268.58 & 6390.34 & 6320.25 \\
 WBIC & \;\;310.10 & \;\;317.60 & \;\;313.36 \\
 Negative ELBO & 3834.33 & 3855.40 & 3840.05 \\
 Strictly Identifiable? & \text{Yes} & \multicolumn{1}{c}{\begin{tabular}{c} \text{No} \\ ($\bm{p}$-partially identifiable) \end{tabular}}& \multicolumn{1}{c}{\begin{tabular}{c} \text{No} \\ ($\bm{p}$-partially identifiable) \end{tabular}} \\
 \midrule
 Time (seconds) & \multicolumn{1}{c}{20} & \multicolumn{1}{c}{1035} & \multicolumn{1}{c}{23} \\
 \bottomrule
 \end{tabular}%
 }
 \begin{tablenotes}
 \item \footnotesize{\textit{Note.} To compute WAIC and WBIC, we run MCMC estimation using four chains with 5,000 iterations and 2,000 burn-in.}
 \end{tablenotes}
\label{table:results_Fraction}
\end{table}%

\subsubsection{ECPE Dataset}
We conducted a comparison among the three Q-matrices estimated from the proposed method, Gibbs sampler, and EM-based algorithm. These Q-matrices are provided in Table B2 of Supplementary Material B. The result (Table \ref{table:results_ECPE}) shows that, in terms of the relative model fit, the Q-matrix from the Gibbs sampler produced the best relative model fit. However, the proposed method also yielded values that were highly comparable to those obtained using the Gibbs sampler. The proposed method outperformed the Q-matrix from the EM-based algorithm for all the five relative model fit indices. The estimated Q-matrices from the three methods all satisfy the strict identifiability for the DINA model. Regarding computation time, the proposed method estimated the Q-matrix approximately 42 times faster than the Gibbs sampler. Although the computation for the EM-based algorithm was approximately two times faster than our method, our method estimated a more optimal Q-matrix than the EM-algorithm in a reasonable time. The estimated Q-matrix from our method coincided with \textcolor{black}{95.24\% and 79.76\%} entries of the Q-matrices from the Gibbs sampler and EM-based algorithm, respectively.

\begin{table}[!htbp]
\color{black}
 \centering
 \caption{ECPE dataset: Relative model fit indices and computation time}
 \resizebox{\linewidth}{!}{
 \begin{tabular}{r CCC}
 \toprule
 & \multicolumn{1}{c}{Proposed method} & \begin{tabular}{c} Gibbs sampler\\ \parencite{chung_gibbs_2019} \end{tabular} & \begin{tabular}{c} EM-based algorithm \\ \parencite{chen_statistical_2015} \end{tabular} \\
 \midrule
 AIC & {85666.38} & 85651.93 & 85791.88 \\
 BIC & {86043.13} & 86028.67 & 86168.63 \\
 WAIC & {83879.7} & 83840.94 & 84156.34 \\
 WBIC & \;\;{3562.28} &\;\;3560.77 & \;\;3588.73 \\
 Negative ELBO &{42993.74} & 42986.46 & 43056.15 \\
 Strictly Identifiable? & \text{Yes}& \text{Yes}& \text{Yes}\\
 \midrule
 Time (seconds) & \multicolumn{1}{c}{86} & \multicolumn{1}{c}{3590} & \multicolumn{1}{c}{42} \\
 \bottomrule
 \end{tabular}%
 }
\begin{tablenotes}
 \item \footnotesize{\textit{Note.} To compute WAIC and WBIC, we run MCMC estimation using four chains with 5,000 iterations and 2,000 burn-in.}
 \end{tablenotes}
 \label{table:results_ECPE}
\end{table}%

\subsubsection{TIMSS 2003 Mathematics Dataset}
The compared Q-matrices are provided in Table B3 of Supplementary Material B. The values of the five relative model fit indices for the two Q-matrices are shown in Table \ref{table:results_TIMMS}. The proposed method estimated the Q-matrix with better relative model fit than the expert-defined Q-matrix. This result corroborates the sound accuracy of our method in light of relative model fit. In addition, although the estimated Q-matrix from our method does not satisfy the strict identifiability, this Q-matrix can still be used to guide practitioners to identify over-specified or irrelevant attributes and to remove such attributes for the smaller attribute's dimensionality. For example, the 10th attribute in the estimated Q-matrix is measured by only one item. Thus, this information can be utilized to remove the 10th attribute and reduce the attribute's dimensionality for better Q-matrix specifications. In terms of computation time, our method took 3 h 13 min for the estimation. However, we can reduce it by leveraging the GPU parallelization for array operation as our method involves the computation of large matrices. The estimated Q-matrix matched 88.29\% entries of the expert-defined matrix. These results verify the utility and scalability of our method under large-scale settings common in DCM's applications.
 
\begin{table}[!htbp]
\color{black}
 \centering
 \caption{TIMSS 2003 mathematics dataset: Relative model fit indices and computation time}
\scalebox{1}{
 \begin{tabular}{r C C}
 \toprule
 & \multicolumn{1}{c}{Proposed method} & \multicolumn{1}{c}{\begin{tabular}{c} Expert knowledge\\ \parencite{su_hierarchical_2013} \end{tabular}} \\
 \midrule
 AIC & 35420.71 &35743.76 \\
 BIC & 73552.78 &73875.83 \\
 WAIC & 18581.60 & 19291.64\\
 WBIC & \quad 803.34 & \quad 855.29\\
 Negative ELBO & 10752.03 & 10929.17\\
 Strictly Identifiable? & \text{No} & \text{No} \\
 \midrule
 Time & \multicolumn{1}{c}{3h 13min} & \multicolumn{1}{c}{$-$}\\
 \bottomrule
 \end{tabular}%
 }
\begin{tablenotes}
 \item \footnotesize{\textit{Note.} To compute WAIC and WBIC, we run MCMC estimation using four chains with 5,000 iterations and 2,000 burn-in.}
 \end{tablenotes}
 \label{table:results_TIMMS}
\end{table}%

\section{Simulation Investigation on Potential Misspecifications Using a Real Dataset}
\subsection{Effects of the Initial Value Choices}
One concern in Q-matrix estimation is the effect of initial value choices. That is, how much stably Q-matrix estimation methods can estimate an optimal Q-matrix from different initial values. As appropriate initial values are unknown in real-world settings, it is crucial to investigate the degree of robustness to misspecified initial values and compare its degree with existing methods. Therefore, we performed the simulation investigation using the fraction subtraction dataset to assess how much stably the proposed method can estimate an optimal Q-matrix from randomly generated initial values and compared its performance with the EM-based algorithm.

Specifically, we estimated a Q-matrix from the fraction subtraction dataset 10,000 times by the proposed method and EM-based algorithm with randomly generated initial values. After all these estimations, we computed the log-likelihood's values given the 10,000 Q-matrices from these two methods. For the proposed method, initial Q-matrices were randomly generated for each estimation. For the EM-based algorithm, we randomly generated the initial values for intercepts and coefficients of attribute main- and interaction-effect terms from the uniform distribution $\mathrm{Uniform}(0,1)$ for each estimation. These initial values satisfied the monotonicity constraint on correct-response probabilities for different mastery profiles of required attributes. The initial values for all the structural parameters were set to be $\frac{1}{2^K}$. Other estimation settings were specified in the same manner as the ones in the previous real data analysis for the fraction subtraction dataset.

\begin{figure}[!htbp]
 \begin{center}
 \includegraphics[width=\linewidth]{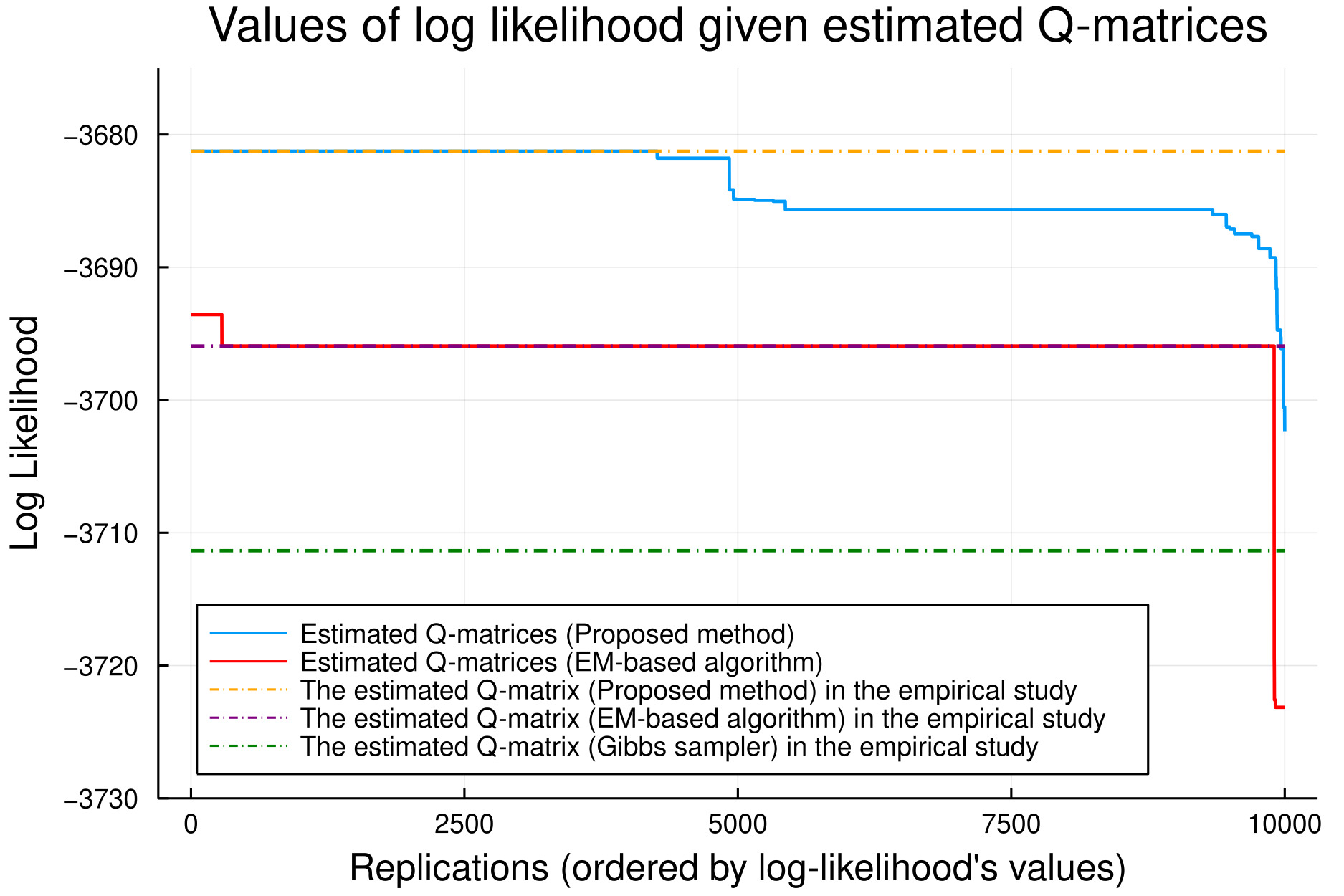}
 \caption{The values of log likelihood given the 10,000 estimated Q-matrices from the proposed method and EM-based algorithm in the simulation investigation using the fraction subtraction dataset. The $x$-axis represents the number of replications, and the $y$-axis represents the log likelihood with all data points given the 10,000 estimated Q-matrices. The blue line corresponds to the log-likelihood's values given the Q-matrices estimated by the proposed method, and the red line corresponds to those of the Q-matrices estimated by the EM-based algorithm. Other dash-dot lines indicate the values of log likelihood given the estimated Q-matrices from the proposed method, EM-based algorithm, and Gibbs sampler in the empirical study.}
\label{fig:EffectsofInitialValues}
 \end{center}
\end{figure}

Figure \ref{fig:EffectsofInitialValues} shows the values of log likelihood given the 10,000 estimated Q-matrices from the proposed method and EM-based algorithm in the simulation investigation. Table \ref{table:EffectsofInitialValues} summarizes the result of this simulation. As evident in the figure and table, the proposed method could stably estimate the identified Q-matrix producing the larger value of log likelihood than the EM-based algorithm. Our method also reached that Q-matrix 4262 times in the 10,000 estimations with randomly generated initial values. Conversely, the EM-based algorithm reached the Q-matrix with its largest value of log likelihood only 282 times. These results corroborate that the proposed method holds the satisfactory degree of robustness to the choice of initial values compared with the existing method.

\begin{table}[!htbp]
\color{black}
\centering
\caption{Summary of the simulation investigation on effects of the initial value choices}
\resizebox{\linewidth}{!}{
 \begin{tabular}{lcc}
 \toprule
& Proposed method & \multicolumn{1}{c}{\begin{tabular}{c} EM-based algorithm \\ \parencite{chen_statistical_2015} \end{tabular} } \\
 \midrule
 \begin{tabular}{l} The maximum value of \\ log likelihood in the replications \end{tabular} & -3681.259 & -3693.565 \\
 \begin{tabular}{l} The number of replications with \\ the maximum value of log likelihood \end{tabular} & 4262& 282 \\
 \begin{tabular}{l} Did the Q-matrix with the maximum value \\ of log likelihood satisfy strict identifiability? \end{tabular} & \text{Yes} & \multicolumn{1}{c}{\begin{tabular}{c} \text{No} \\ ($\bm{p}$-partially identifiable) \end{tabular}} \\
 \bottomrule
 \end{tabular}%
 }
\label{table:EffectsofInitialValues}%
\end{table}%

\subsection{Effects of Non-optimal Hyperparameters}
Another concern related to the proposed method is the specification of hyperparameters: the number of times to run the iteration algorithm $R$, number of iterations in a single run $T$, number of discarded iterations for iterative averaging $D$, and \textcolor{black}{mini-batch} sample size for the stochastic optimization $S$. To investigate the effects of non-optimal hyperparameters for the Q-matrix estimation and identify the one that has the most salient impact on the optimality of the estimation, we designed another simulation investigation using the fraction subtraction dataset. Note that this simulation does not include the number of run $R$ as an experimental factor and sets it to $R=1$ for all the following conditions. This is because the effect of other hyperparameters can be mitigated by initial Q-matrices across multiple runs that happen to be close to the optimal one. In such a case, this factor causes confounding for evaluating the effects of other hyperparameters.

This simulation considers the four conditions: the control, $T\;\textit{decreased}$, $D\;\textit{decreased}$, and $S\;\textit{decreased}$ conditions. For the control condition, we set $T=550$, $D=50$, and $S=300$, which are the same hyperparameter specification in the empirical study for the fraction subtraction dataset. Then, we reduced $T=550$ to $T=300$, $D=50$ to $D=0$, and $S=300$ to $S=150$ for the $T\;\textit{decreased}$, $D\;\textit{decreased}$, and $S\;\textit{decreased}$ conditions, respectively. The estimation was conducted 10,000 times under the four conditions, and we computed the values of ELBO given the estimated Q-matrices from these conditions.

\begin{figure}[!htbp]
\color{black}
 \begin{center}
 \includegraphics[width=\linewidth]{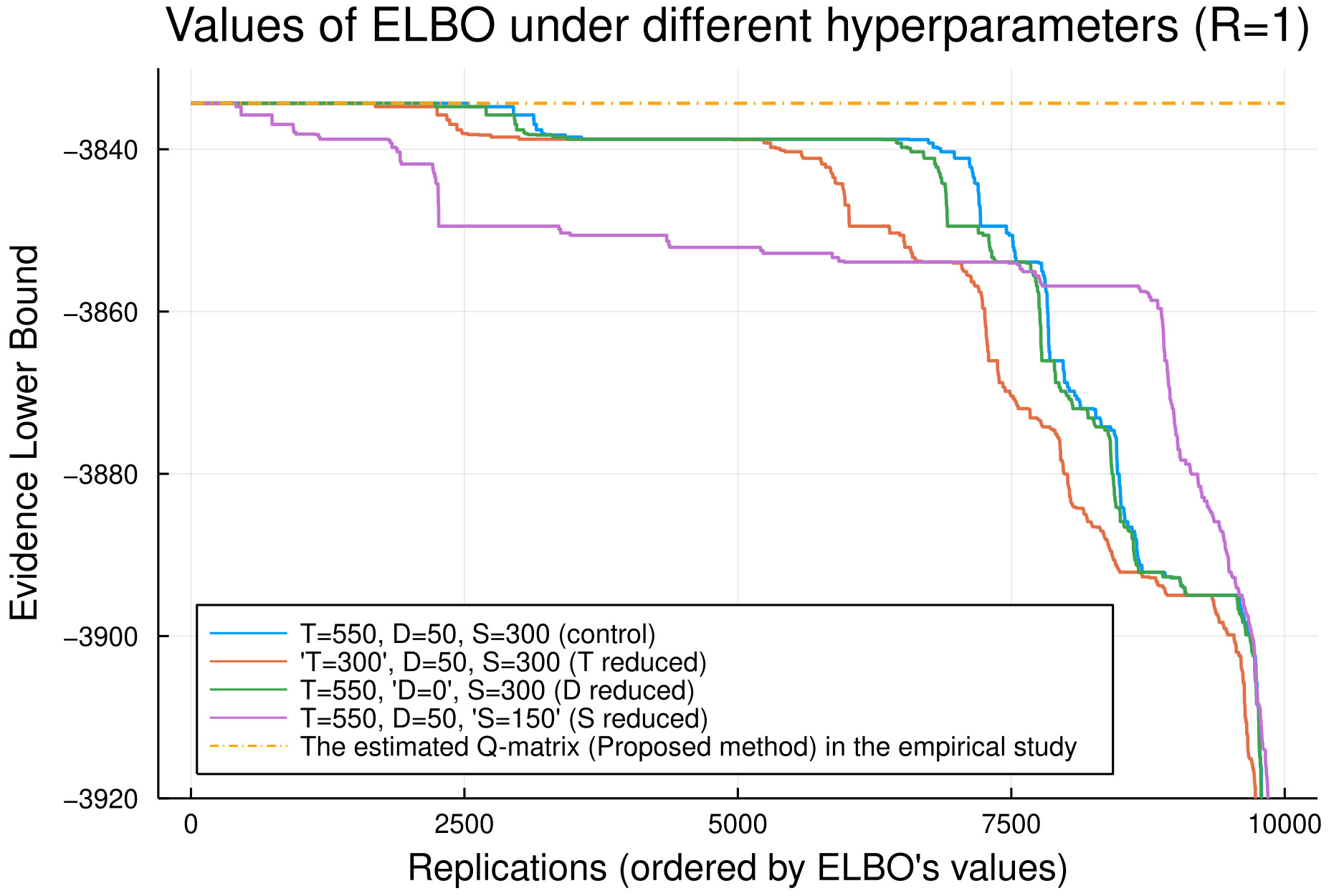}
 \caption{Values of ELBO given the 10,000 estimated Q-matrices from the proposed method under the four conditions in the simulation investigation using the fraction subtraction dataset. The $x$-axis represents the number of replications, and the $y$-axis represents the ELBO with all data points given the 10,000 estimated Q-matrices. Each colored line corresponds to the ELBO's values given the 10,000 estimated Q-matrices under the conditions. The yellow dash-dot line indicates the value of ELBO given the estimated Q-matrix from the proposed method in the empirical study.}
\label{fig:Effectsofhyperparameters}
 \end{center}
\end{figure}

Figure \ref{fig:Effectsofhyperparameters} shows the values of ELBO given the 10,000 estimated Q-matrices from the proposed method under the four conditions. Table \ref{table:Effectsofhyperparameters} summarizes the result of this simulation. We observed that the most impactful factor among the three hyperparameters is the number of \textcolor{black}{mini-batch samples} for the stochastic optimization $S$. Although 2537 estimated Q-matrices reached the most optimal solution in the control condition, only 412 estimated Q-matrices reached that point when $S$ was reduced to $150$. This indicates that the misspecification on the size of $S$ deteriorates the stability of Q-matrix estimation. In terms of other hyperparameters, the number of iterations in a single run $T$ had a moderate effect on the estimation stability. Here, the number of the estimated Q-matrices reaching the most optimal solution was 1686. The number of discarded iterations for iterative averaging $D$ presented a small effect on the estimation stability. Although the non-optimal hyperparameters degraded the stability of estimating an optimal Q-matrix to some extent, the estimation under all the conditions reached the most optimal solution. These results suggest that non-optimal hyperparameters can deteriorate the estimation stability; however, the proposed method is robust to these hyperparameters in terms of the optimality of Q-matrix estimation. It should be noted that all the replications across the four conditions were implemented with $R=1$ so that these negative effects would be less salient under the larger size of $R$.

\begin{figure}[!htbp]
\color{black}
 \begin{center}
 \includegraphics[width=\linewidth]{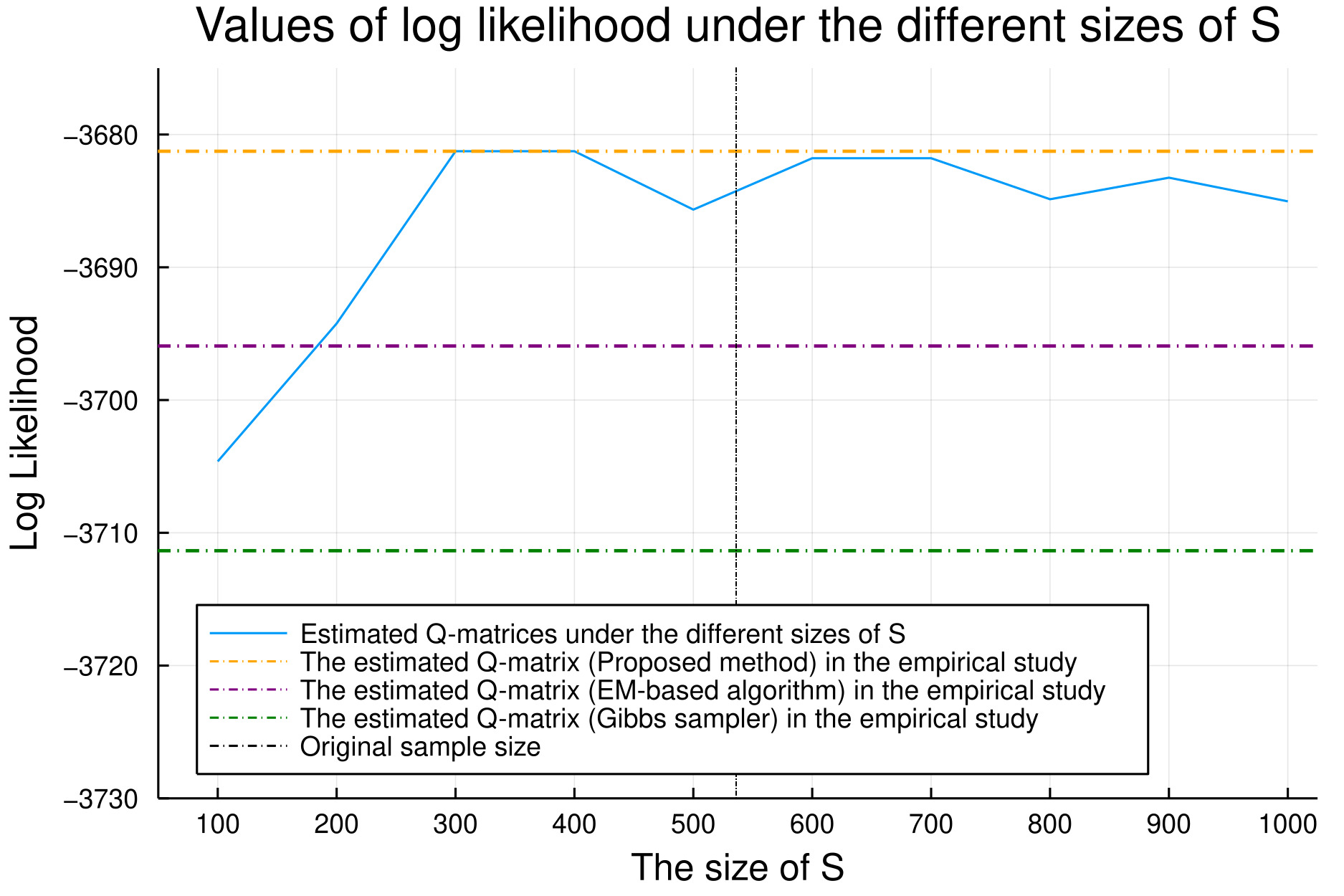}
 \caption{Values of log likelihood given the estimated Q-matrices from the proposed method under the different sizes of $S$ in the simulation investigation using the fraction subtraction dataset. The $x$-axis represents the size of $S$, and the $y$-axis represents the log likelihood with all data points given the estimated Q-matrices. The vertical black dot line represents the original sample size of the fraction subtraction dataset. The blue line corresponds to the log-likelihood's values given the estimated Q-matrices under the different sizes of $S$. Other dash-dot lines indicate the values of log likelihood given the estimated Q-matrices from the proposed method, EM-based algorithm, and Gibbs sampler in the empirical study.}
\label{fig:multipleS}
 \end{center}
\end{figure}

Finally, we conducted a further investigation on the effect of the \textcolor{black}{mini-batch} sample size $S$. To evaluate how the different sizes of $S$ make the estimation unstable, we increased the size of $S$ by $100$ from $100$ to $1000$, estimated a Q-matrix using the fraction subtraction dataset under these 10 different sizes of $S$, and computed the log-likelihood's values with the ten estimated Q-matrices. For all the estimations, we set $R=10$, $T=550$, and $D=50$. Figure \ref{fig:multipleS} presents the results. These results indicate that the proposed method stably estimated a more optimal Q-matrix than the EM-algorithm when $S$ was more than $200$, and this superiority holds over the Gibbs sampler across all the settings of $S$. The additional investigation supports the satisfactory degree of robustness to the potential misspecification of the most influential hyperparameter $S$ 
in the proposed method.

\begin{table}[!htbp]
\color{black}
\centering
\caption{Summary of the simulation investigation on effects of hyperparameters}
\resizebox{\linewidth}{!}{
 \begin{tabular}{lCCCC}
 \toprule
& \text{Control} & $T\;\textit{decreased}$ & $D\;\textit{decreased}$ & $S\;\textit{decreased}$\\
 \midrule
 \begin{tabular}{l} Maximum value of \\ ELBO in the replications \end{tabular} & -3834.33 &-3834.33 &-3834.33 &-3834.33 \\
 \begin{tabular}{l} Number of replications with \\ the maximum value of ELBO \end{tabular} & 2537 & 1686 & 2235 & 412 \\
 \begin{tabular}{l} Did the Q-matrix with the maximum value \\ of ELBO satisfy strict identifiability? \end{tabular} & \text{Yes} & \text{Yes} & \text{Yes} & \text{Yes} \\
 \bottomrule
 \end{tabular}%
 }
\label{table:Effectsofhyperparameters}%
\end{table}%

\section{Discussion and Conclusion}
This study set out to develop a scalable Bayesian Q-matrix estimation algorithm for the DINA Q-matrix. In the proposed framework, we introduced a new framing of Q-matrix estimation, where the estimation is regarded as a root-finding problem in which the goal is to find the Q-matrix with the maximum marginal likelihood. Then, we constructed an iteration algorithm for optimizing a Q-matrix utilizing stochastic optimization and variational inference. These two techniques achieve high scalability and effective optimization, making it possible to estimate a Q-matrix in a fast and accurate manner. 

To assess the performance of the proposed method, we conducted a simulation study under the small- and large-scale conditions. The findings from the simulation are that most of the element-wise mean recovery rates were more than 93\% in small-scale conditions and approximately 97\% in large-scale conditions. In some small-scale conditions (e.g., $K=4$, $J=10$, and $N=250$ under $\rho=0$ and $K=4$, $J=10$, and $N=250$ and $500$ under $\rho=0.25$), the element-wise mean recovery rate were 85--90\%. However, when the sample size increased to $1000$, these rates became over 93\%. The full recovery of the true Q-matrices was generally satisfactory in small-scale conditions especially with $K=3$. This became more difficult in large-scale conditions because the number of elements in a Q-matrix is much larger than the small size of a Q-matrix. However, when the sample size increased to $8000$, the proposed method could estimate all the entries of the true Q-matrices from one-fifth to \textcolor{black}{one-half} of the replications. Overall, the proposed method exhibited good accuracy under a wider range of conditions.

Through applications to real datasets, we found that the estimated Q-matrix from the proposed method outperforms other Q-matrices estimated by the Gibbs sampler and EM-based algorithm in terms of relative model fit indices, except for the analysis of relative model fit in the ECPE dataset. Although the superiority of the Gibbs sampler over the proposed method in terms of relative model fit was observed in the ECPE dataset, our method produced values that were highly comparable to those of the Gibbs sampler. In addition, the computation time of our method is approximately 42-52 times faster than the Gibbs sampler, and our method would benefit from its computational advantage even more as the number of attributes increases. The computation time was also comparable with that of the EM-based algorithm. These results provide empirical evidence on the reliable accuracy and fast computation of the proposed method. Furthermore, we estimated a Q-matrix with highly dimensional attributes of $K=13$, and the estimated Q-matrix fit the data better than the expert-defined Q-matrix. Although this estimation took approximately 3 hours, we would be able to curtail it significantly by implementing the GPU parallelization for array operation because matrix computation under a large $K$ involves large matrices, which presents a bottleneck for fast computation. 

\textcolor{black}{Regarding the identifiability issue of Q-matrix estimation, estimating an identified Q-matrix commonly becomes infeasible in practice, such as in a situation similar to the TIMSS 2003 mathematics dataset. However, our method can help practitioners identify over-specified or irrelevant attributes for better Q-matrix specifications because this method can estimate an optimal Q-matrix from the vast space comprising all types of possible Q-matrices in a feasible time. Besides, we conducted the additional simulation study to compare our method with other estimation methods equipped with identification constraints and revealed that our method is highly comparable with these methods (Supplementary Material 
D). This result shows that our method estimated an identified, true Q-matrix in an accurate and fast manner from a much greater Q-matrix space than the space being considered by the methods with identification constraints. Hence, the lack of identification constraints does not hinder our method from estimating an identified, true Q-matrix as accurately as the methods with such constraints. Furthermore, we empirically found that marginal likelihood, whose lower bound (ELBO) is adopted in the proposed method, is able to differentiate an unidentified, true Q-matrix from other Q-matrices that have the same log-likelihood value as the true one. For instance, one unidentified, true Q-matrix with $K=4$ and $J=6$ provided the log-likelihood value of -346630.739 under the DINA model and $N=100,000$. However, another Q-matrix also yielded the same log-likelihood value as the true one. In contrast, the ELBO value of that true Q-matrix differed from that of another one, where the former value was -346763.324, whereas the latter one was -346765.463. This empirically shows that the ELBO successfully selects the true Q-matrix despite the fact that the log likelihood failed to differentiate it from other Q-matrices. This behavior of ELBO is highly desirable because it indicates that Q-matrix estimation based on marginal likelihood is capable of finding the optimal Q-matrix even when other Q-matrices produce the same log-likelihood value as that of the unidentified, true Q-matrix. This example's details and other examples are provided in Supplementary Material E. Finally, as \textcite{sessoms_applications_2018} reported in their DCM's literature review (i.e., the average number of attributes adopted in DCM applications is eight), the situation where estimating an identified Q-matrix is infeasible can frequently occur in application settings. In such a case, our method can be a useful data-driven tool for practitioners to develop a Q-matrix.}

The following are the two limitations of this study. First, the extension of the proposed algorithm to other DCMs than the DINA model is desirable. As the deterministic inputs, noisy ``or'' gate \parencite[DINO;][]{templin_measurement_2006} model differs from the DINA model only in its form of ideal response, the proposed algorithm can be readily extended to the DINO Q-matrix estimation. Further, our method can be extended to the saturated case by adopting the new algorithm for variational inference of the saturated DCM \parencite{yamaguchi_variational_2021}, although modifying the selection procedure of an updated Q-matrix is required. This topic is left to future research. From the additional analysis on the TIMSS 2003 mathematics dataset using AIC and BIC (Table C1 of the Supplementary Material C), however, we observed that the estimated Q-matrix from our method yielded better relative model fit than the expert-defined Q-matrix even in other DCMs, such as DINO, RRUM, ACDM, and GDINA, with respect to AIC and in DINO, ACDM, and GDINA with respect to BIC. This result indicates that a good specification of the Q-matrix may be preserved across different models. If this is the case, it is conceivable that the Q-matrix obtained from our method can be utilized as a data-driven guide for Q-matrix development even under different model assumptions. Second, the number of attributes was assumed to be known in this study. Future research can work on a method to estimate the attribute dimensionality and Q-matrix simultaneously. Techniques to determine the number of clusters in finite mixture models can be applied to estimate the number of attributes because DCMs can be reframed within a class of mixture models.

Finally, potential misspecifications of initial values and hyperparameters on the proposed method should be considered before its implementation. From the simulation investigation using the fraction subtraction dataset, we revealed that our method holds sound robustness toward the choice of initial Q-matrices and non-optimal hyperparameters. For better stability of the Q-matrix estimation, we recommend using an expert-defined Q-matrix as an initial value if it is available. For hyperparameters, the number of \textcolor{black}{mini-batch samples} for the stochastic optimization $S$ can influence the stability of the estimation. We recommend based on the simulation and empirical evidence, that the size of $S$ should be more than $300$. This setting would suffice for a stable and accurate estimation. When this size is set to too small such as $100$, excessive random noise induced by sampling variations hinders the updated Q-matrices from converging to the neighborhood of an optimal Q-matrix. For other hyperparameter specifications, the decrease in the number of iterations $T$ moderately deteriorates the estimation stability. Hence, we recommend that $T$ be more than $500$. The number of discarded iterations $D$ has an ignorable effect on the estimation. However, the increase in $D$ can ensure that an estimated Q-matrix is computed from updated Q-matrices in the neighborhood of an optimal Q-matrix, especially when the size of a Q-matrix is large. \textcolor{black}{It should be noted that, although we empirically showed the sound robustness of our method toward hyperparameter specifications, it is desirable to conduct multiple trials with different hyperparameter values (e.g., different mini-batch samples) and select the one with the largest ELBO if computational capacity is available because such a hyperparameter tuning is a common practice in data analysis to avoid unsuitable hyperparameter values.}

In summary, the proposed scalable method for Bayesian Q-matrix estimation offers a tool for the data-driven development of a Q-matrix in large-scale settings. As noted by \textcite{yamaguchi_variational_2020}, large-scale applications in DCMs have recently made inroads \parencite[for example,][]{chen_procedure_2014,jang_improving_2019,yamaguchi_comparison_2018}. Thus, a rise in the demand for a scalable Q-matrix estimation method can be expected. The proposed method would contribute to meeting such demand and can help practitioners construct a Q-matrix in these settings.

\section*{Conflict of Interest}
The authors declare that they have no conflict of interest.

\section*{Acknowledgment}
We would like to thank Dr. Yunxiao Chen for sharing the code for the EM-based Q-matrix estimation algorithm with Lasso regularization used in the empirical study. This work was supported by JSPS KAKENHI Grant Number 17H04787, 21H00936 and JST, PRESTO Grant Number JPMJPR21C3, Japan. Preliminary reports of this study were presented at the 48th annual meeting of the Behaviormetric Society of Japan and the International Meeting of Psychometric Society (IMPS) 2021.

\printbibliography

\clearpage
\section*{Supplementary Materials}
\subsection*{Supplementary Material A}
The following are the true Q-matrices for the simulation study. Regarding the small-scale conditions, the three-attribute Q-matrix with 10 items comprises two sets of the identity matrix and all possible $q$-vectors that require two and three attributes. We then concatenated two of the three-attribute Q-matrix for the Q-matrix with 20 items. Specifically, the four-attribute Q-matrix with 10 items contains one set of the identity matrix, all possible $q$-vectors that require two attributes, and two $q$-vectors that measure three attributes. Meanwhile, the four-attribute Q-matrix with 20 items includes two sets of the identity matrix, two sets of all possible $q$-vectors that require two attributes, and four $q$-vectors that measure three attributes.

Regarding the large-scale conditions, the seven-attribute Q-matrix with 40 items comprises two sets of the identity matrix and 26 randomly selected items from the set of possible $q$-vectors that require two and three attributes. The 26 items were randomly selected in such a manner that half of those $q$-vectors measure two attributes, and the other half measure three attributes. In addition, the number of items that measure each attribute was set to be approximately equal across attributes. The seven-attribute Q-matrix with 80 items is the double stacking of the seven-attribute Q-matrix with 40 items. The eight-attribute Q-matrix contains two sets of the identity matrix and 24 randomly selected items from the set of possible $q$-vectors that require two and three attributes. The 24 items were randomly selected in the same manner as the seven-attribute Q-matrix with 40 items. Additionally, the eight-attribute Q-matrix with 80 items is the double stacking of the eight-attribute Q-matrix with 40 items.

In the following figures for the specifications of the true Q-matrices, a white box denotes an entry of a Q-matrix that takes the value of 1, and a black box denotes the one that takes the value of 0. The files for these true Q-matrices can be obtained from the data repository in the Open Science Framework: \url{https://osf.io/jev9q/?view_only=e1b1f047c89f46cba9a3b61194404d8e}.
\clearpage
\begin{figure}[!htbp]
 \begin{center}
 %\begin{tabular}{c}
 \begin{minipage}{1\hsize}
 \centering
 \includegraphics[width=1\linewidth]{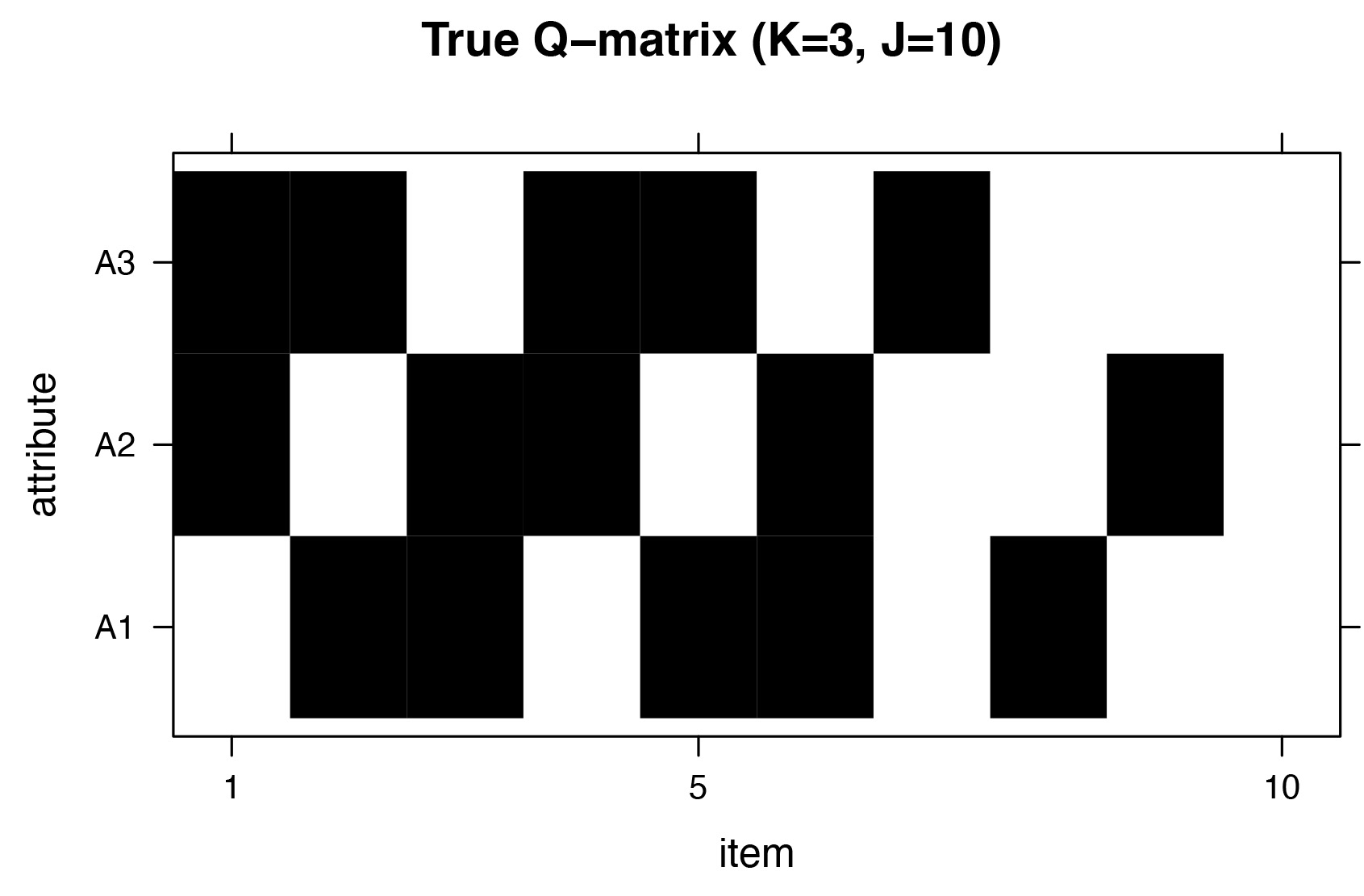}
 \end{minipage} 
 \begin{minipage}{1\hsize}
 \centering
 \includegraphics[width=1\linewidth]{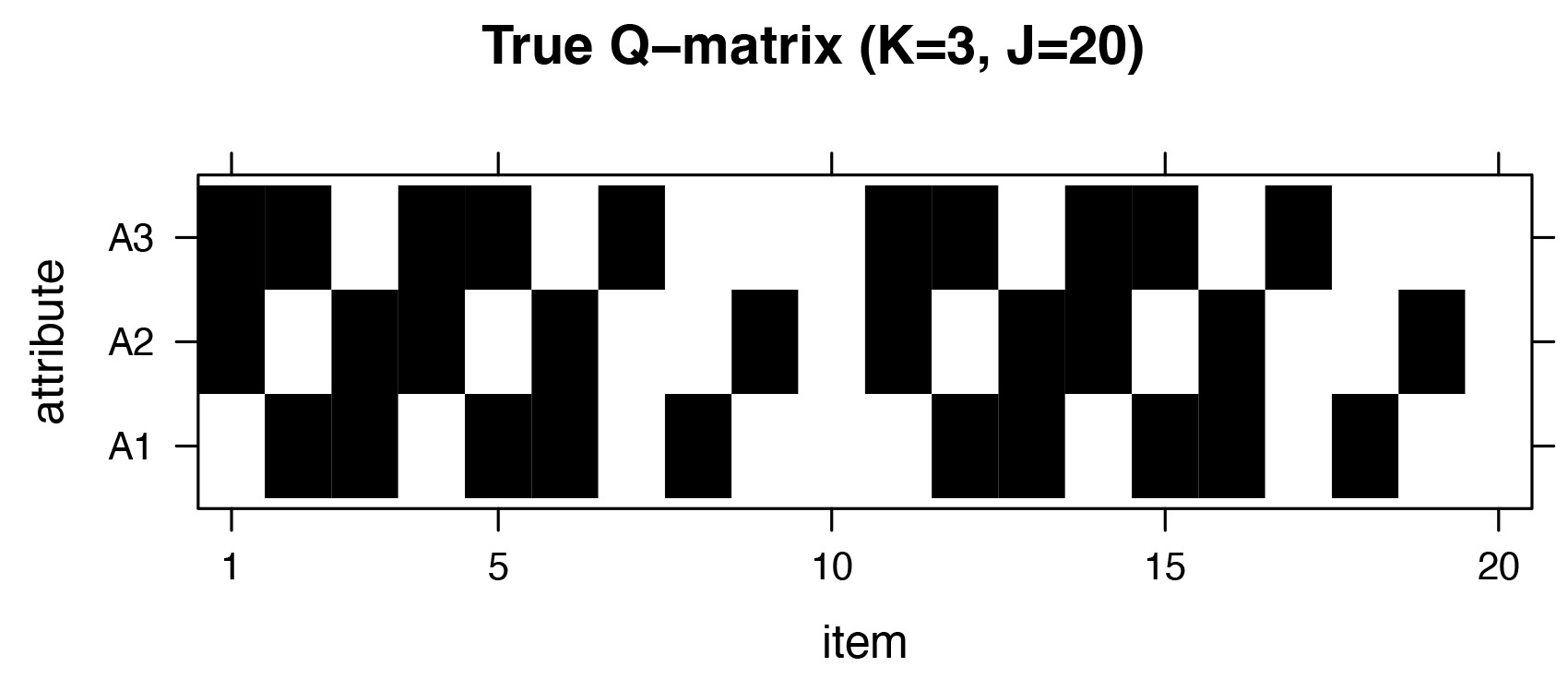}
 \end{minipage}
 %\end{tabular}
 \end{center}
\end{figure}

\begin{figure}[!htbp]
 \begin{center}
 %\begin{tabular}{c}
 \begin{minipage}{1\hsize}
 \centering
 \includegraphics[width=1\linewidth]{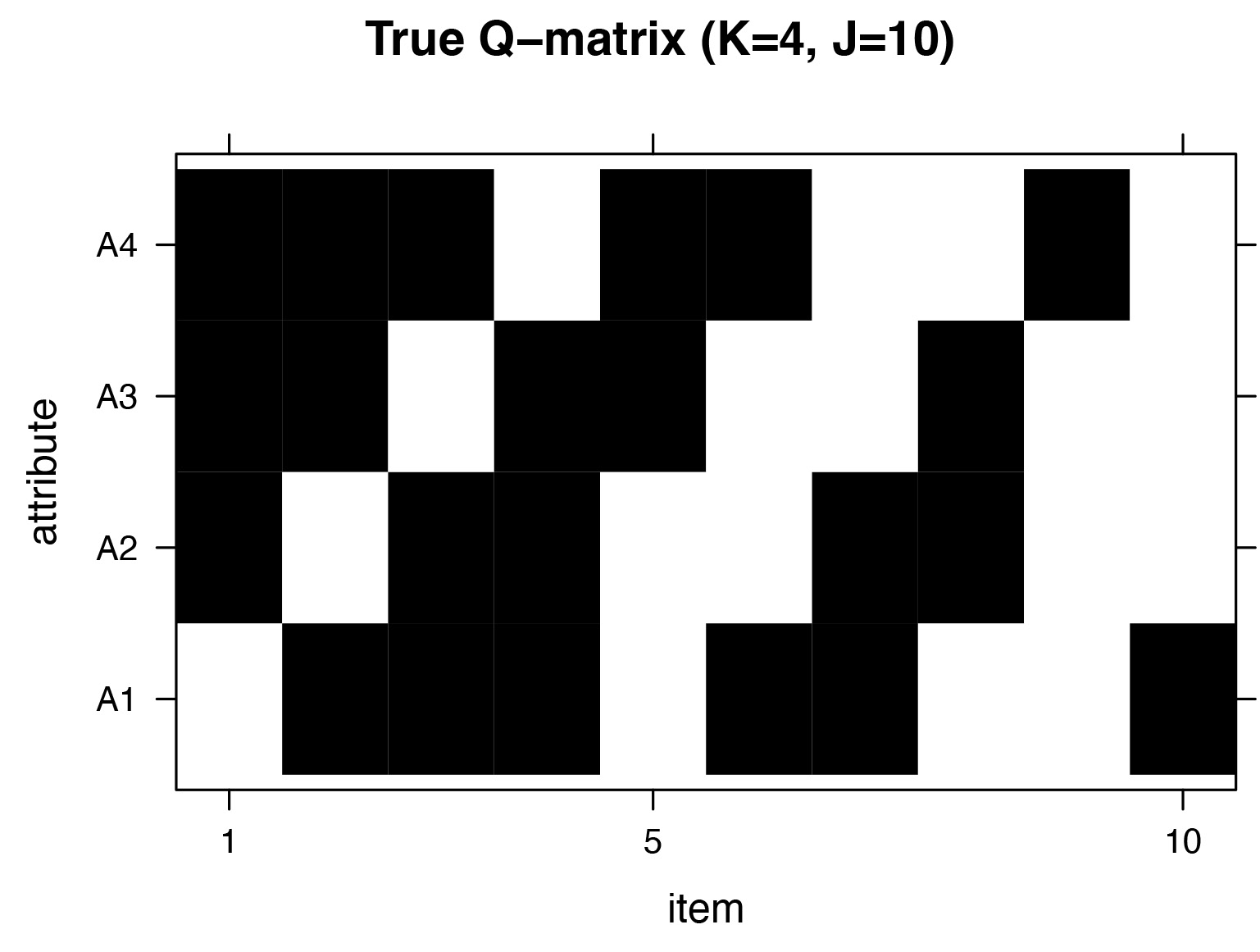}
 \end{minipage} 
 \begin{minipage}{1\hsize}
 \centering
 \includegraphics[width=1\linewidth]{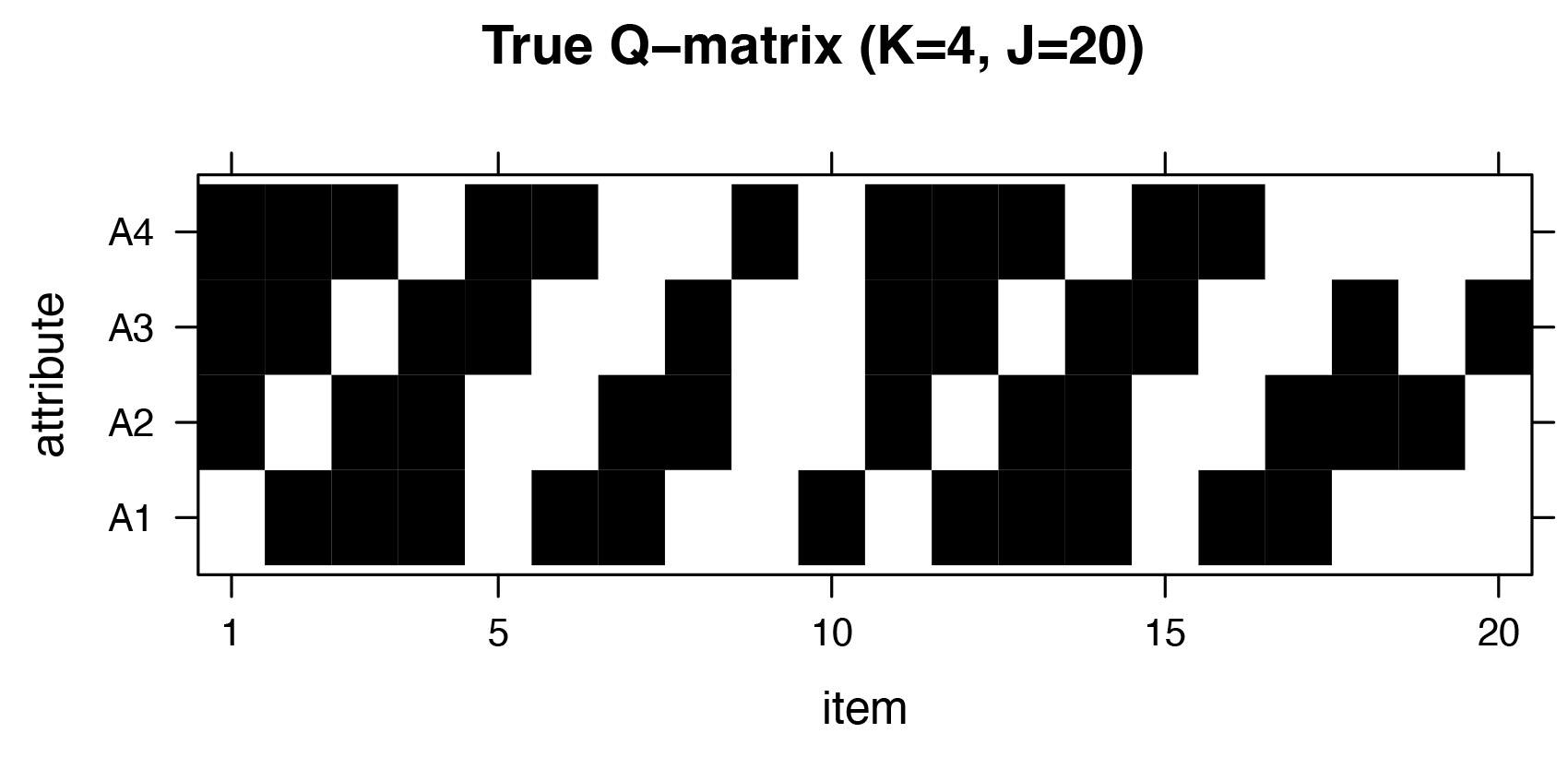}
 \end{minipage}
 %\end{tabular}
 \end{center}
\end{figure}

\begin{landscape}
\begin{figure}[!htbp]
 \includegraphics[width=1\linewidth]{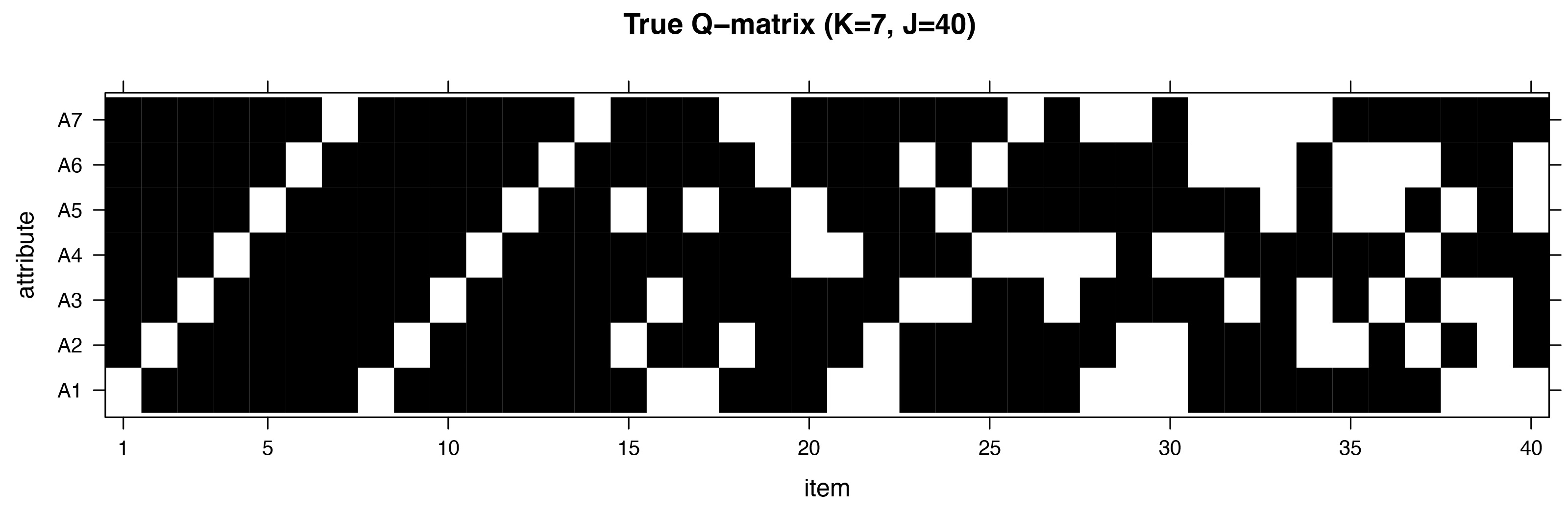}
\end{figure}
\begin{figure}[!htbp]
 \includegraphics[width=1\linewidth]{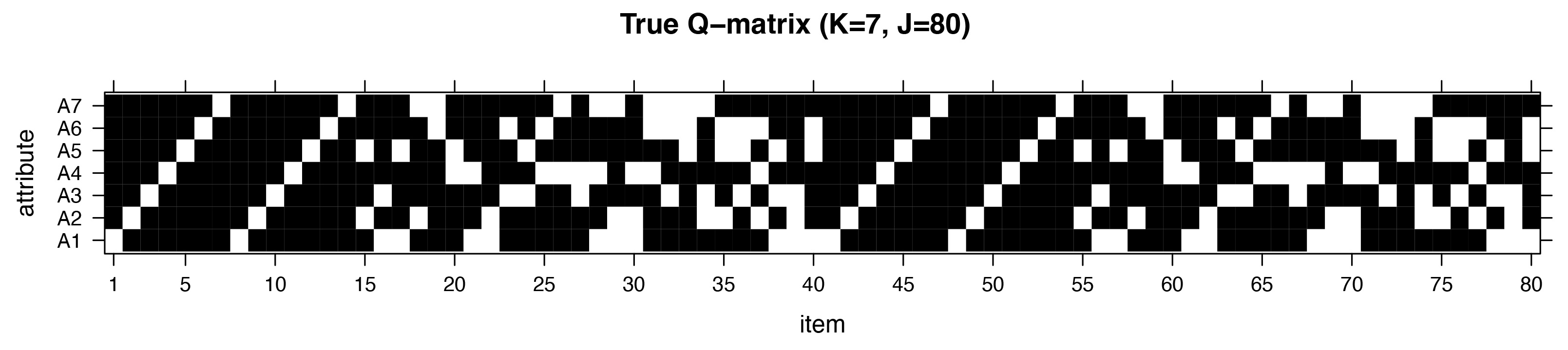}
\end{figure}
\end{landscape}

\begin{landscape}
\begin{figure}[!htbp]
 \includegraphics[width=1\linewidth]{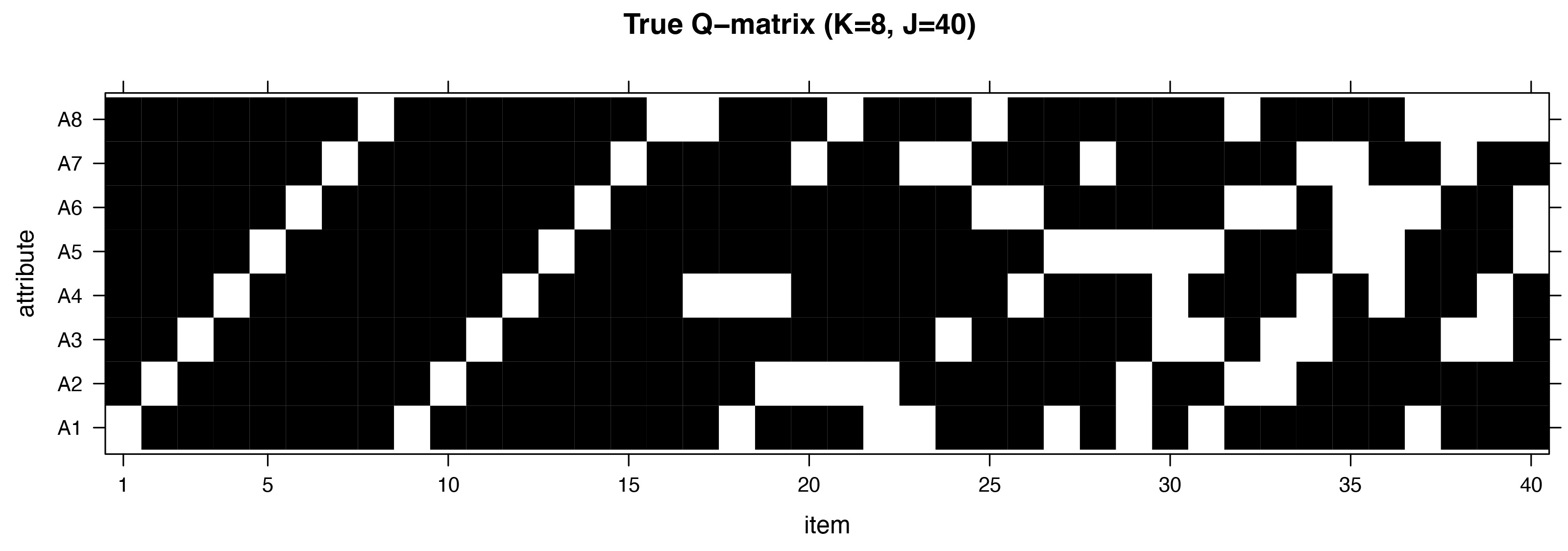}
\end{figure}
\begin{figure}[!htbp]
 \includegraphics[width=1\linewidth]{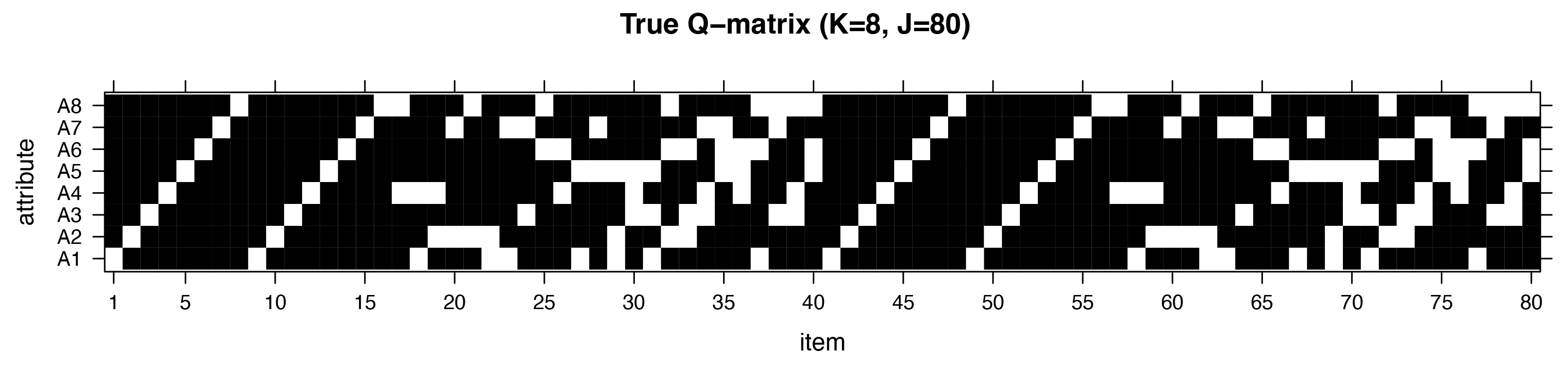}
\end{figure}
\end{landscape}

\clearpage

\subsection*{Supplementary Material B}
\begin{table}[!htbp]
\renewcommand\thetable{B1}
\caption{Q-matrices for the comparison in the fraction subtraction dataset}
\centering
\resizebox{\linewidth}{!}{
\begin{tabular}{ccccccccccccccccc}
\toprule
\multicolumn{5}{c}{Proposed method} & & \multicolumn{5}{c}{\begin{tabular}{c} Gibbs sampler\\ \parencite{chung_gibbs_2019}\end{tabular}} & & \multicolumn{5}{c}{ \begin{tabular}{c} EM-based algorithm \\ \parencite{chen_statistical_2015} \end{tabular}} \\
\cmidrule{1-5}\cmidrule{7-11}\cmidrule{13-17}\multirow{2}[4]{*}{Item} & \multicolumn{4}{c}{Attributes} & & \multirow{2}[4]{*}{Item} & \multicolumn{4}{c}{Attributes} & & \multirow{2}[4]{*}{Item} & \multicolumn{4}{c}{Attributes} \\
\cmidrule{2-5}\cmidrule{8-11}\cmidrule{14-17}& A1& A2& A3& A4& & & A1& A2& A3& A4& & & A1& A2& A3& A4 \\
\cmidrule{1-5}\cmidrule{7-11}\cmidrule{13-17}1 & 1 & 0 & 0 & 0 & & 1 & 1 & 0 & 0 & 0 & & 1 & 1 & 0 & 0 & 0 \\
2 & 1 & 0 & 0 & 0 & & 2 & 1 & 0 & 0 & 0 & & 2 & 1 & 0 & 0 & 0 \\
3 & 1 & 0 & 0 & 0 & & 3 & 1 & 0 & 0 & 0 & & 3 & 1 & 0 & 0 & 0 \\
4 & 0 & 1 & 0 & 0 & & 4 & 0 & 1 & 1 & 0 & & 4 & 0 & 1 & 1 & 0 \\
5 & 0 & 0 & 1 & 0 & & 5 & 0 & 0 & 1 & 0 & & 5 & 0 & 0 & 1 & 0 \\
6 & 0 & 0 & 1 & 0 & & 6 & 0 & 0 & 1 & 0 & & 6 & 0 & 0 & 1 & 0 \\
7 & 0 & 0 & 1 & 0 & & 7 & 0 & 0 & 1 & 0 & & 7 & 0 & 0 & 1 & 0 \\
8 & 1 & 0 & 1 & 0 & & 8 & 1 & 0 & 1 & 0 & & 8 & 1 & 0 & 1 & 0 \\
9 & 0 & 0 & 0 & 1 & & 9 & 0 & 0 & 1 & 1 & & 9 & 0 & 0 & 1 & 1 \\
10& 0 & 0 & 1 & 1 & & 10& 0 & 0 & 1 & 1 & & 10& 0 & 0 & 1 & 1 \\
11& 0 & 1 & 0 & 0 & & 11& 0 & 1 & 1 & 0 & & 11& 0 & 1 & 1 & 0 \\
12& 0 & 1 & 1 & 0 & & 12& 0 & 1 & 1 & 0 & & 12& 0 & 1 & 1 & 0 \\
13& 0 & 1 & 1 & 0 & & 13& 0 & 1 & 1 & 0 & & 13& 0 & 1 & 1 & 0 \\
14& 0 & 1 & 1 & 0 & & 14& 0 & 1 & 1 & 0 & & 14& 0 & 1 & 1 & 0 \\
15& 0 & 1 & 1 & 1 & & 15& 0 & 1 & 1 & 1 & & 15& 0 & 1 & 1 & 1 \\
16& 0 & 1 & 1 & 0 & & 16& 0 & 1 & 1 & 0 & & 16& 0 & 1 & 1 & 0 \\
17& 1 & 1 & 1 & 0 & & 17& 1 & 1 & 1 & 1 & & 17& 1 & 1 & 1 & 0 \\
\bottomrule
\end{tabular}%
}
\end{table}%
\clearpage

\begin{landscape}
\begin{table}[!htbp]
\color{black}
\renewcommand\thetable{B2} 
\setlength{\tabcolsep}{12pt}
\centering
\caption{The Q-matrices for the comparison in the ECPE dataset}
\scalebox{0.85}{
\begin{tabular}{cccccccccccccc}
\toprule
\multicolumn{4}{c}{Proposed method} & & \multicolumn{4}{c}{\begin{tabular}{c} Gibbs sampler\\ \parencite{chung_gibbs_2019} \end{tabular}} & & \multicolumn{4}{c}{\begin{tabular}{c} EM-based algorithm \\ \parencite{chen_statistical_2015}  \end{tabular}} \\
\cmidrule{1-4}\cmidrule{6-9}\cmidrule{11-14} 
\multirow{2}[4]{*}{Item} & \multicolumn{3}{c}{Attributes} & & \multirow{2}[4]{*}{Item} & \multicolumn{3}{c}{Attributes} & & \multirow{2}[4]{*}{Item} & \multicolumn{3}{c}{Attributes} \\
\cmidrule{2-4}\cmidrule{7-9}\cmidrule{12-14}& A1& A2& A3& & & A1& A2& A3& & & A1& A2& A3 \\
\cmidrule{1-4}\cmidrule{6-9}\cmidrule{11-14}1 & 1 & 1 & 0 & & 1 & 1 & 1 & 0 & & 1 & 1 & 1 & 0 \\
2 & 0 & 1 & 0 & & 2 & 0 & 1 & 0 & & 2 & 0 & 1 & 0 \\
3 & 1 & 1 & 0 & & 3 & 1 & 1 & 1 & & 3 & 1 & 0 & 1 \\
4 & 0 & 0 & 1 & & 4 & 0 & 0 & 1 & & 4 & 0 & 0 & 1 \\
5 & 0 & 0 & 1 & & 5 & 0 & 0 & 1 & & 5 & 0 & 0 & 1 \\
6 & 0 & 0 & 1 & & 6 & 0 & 0 & 1 & & 6 & 0 & 0 & 1 \\
7 & 1 & 1 & 0 & & 7 & 1 & 1 & 0 & & 7 & 1 & 0 & 1 \\
8 & 0 & 1 & 0 & & 8 & 0 & 1 & 0 & & 8 & 0 & 1 & 0 \\
9 & 1 & 0 & 1 & & 9 & 1 & 0 & 1 & & 9 & 1 & 0 & 0 \\
10& 1 & 0 & 0 & & 10& 1 & 0 & 0 & & 10& 1 & 0 & 0 \\
11& 1 & 0 & 1 & & 11& 1 & 0 & 1 & & 11& 1 & 0 & 1 \\
12& 1 & 0 & 1 & & 12& 1 & 0 & 1 & & 12& 1 & 0 & 1 \\
13& 1 & 0 & 0 & & 13& 1 & 0 & 0 & & 13& 1 & 0 & 0 \\
14& 1 & 1 & 1 & & 14& 1 & 1 & 1 & & 14& 1 & 0 & 0 \\
15& 0 & 1 & 0 & & 15& 0 & 1 & 0 & & 15& 0 & 0 & 1 \\
16& 1 & 0 & 0 & & 16& 1 & 1 & 0 & & 16& 0 & 0 & 1 \\
17& 0 & 1 & 0 & & 17& 0 & 1 & 0 & & 17& 0 & 1 & 1 \\
18& 0 & 1 & 0 & & 18& 0 & 1 & 0 & & 18& 0 & 0 & 1 \\
19& 0 & 0 & 1 & & 19& 0 & 0 & 1 & & 19& 0 & 0 & 1 \\
20& 1 & 0 & 1 & & 20& 1 & 0 & 1 & & 20& 1 & 0 & 1 \\
21& 0 & 1 & 0 & & 21& 0 & 1 & 0 & & 21& 0 & 1 & 0 \\
22& 0 & 0 & 1 & & 22& 0 & 0 & 1 & & 22& 0 & 0 & 1 \\
23& 0 & 1 & 0 & & 23& 0 & 1 & 0 & & 23& 0 & 1 & 0 \\
24& 0 & 1 & 1 & & 24& 0 & 1 & 1 & & 24& 1 & 1 & 1 \\
25& 1 & 1 & 0 & & 25& 1 & 1 & 1 & & 25& 1 & 0 & 0 \\
26& 0 & 1 & 1 & & 26& 0 & 0 & 1 & & 26& 0 & 0 & 1 \\
27& 1 & 0 & 0 & & 27& 1 & 0 & 0 & & 27& 1 & 0 & 0 \\
28& 0 & 0 & 1 & & 28& 0 & 0 & 1 & & 28& 0 & 0 & 1 \\
\bottomrule
\end{tabular}%
}
\end{table}%
\end{landscape}
\clearpage

\begin{table}[!htbp]
\renewcommand\thetable{B3} 
\centering
\caption{The estimated Q-matrix from the TIMSS 2003 mathematics dataset}
\resizebox{\linewidth}{!}{
\begin{tabular}{cccccccccccccc}
\toprule
\multicolumn{14}{l}{Proposed method} \\
\midrule
\multirow{2}[4]{*}{Item} & \multicolumn{13}{c}{Attributes} \\
\cmidrule{2-14}& A1& A2& A3& A4& A5& A6& A7& A8& A9& A10 & A11 & A12 & A13 \\
\midrule
1 & 1 & 0 & 0 & 0 & 0 & 0 & 0 & 0 & 0 & 0 & 1 & 0 & 1 \\
2 & 1 & 0 & 0 & 0 & 0 & 0 & 0 & 0 & 0 & 0 & 0 & 0 & 0 \\
3 & 0 & 1 & 0 & 0 & 0 & 0 & 1 & 0 & 0 & 0 & 0 & 0 & 0 \\
4 & 0 & 0 & 0 & 0 & 0 & 0 & 1 & 0 & 1 & 0 & 1 & 0 & 0 \\
5 & 0 & 0 & 0 & 0 & 0 & 1 & 0 & 0 & 0 & 1 & 0 & 1 & 0 \\
6 & 0 & 0 & 0 & 0 & 0 & 1 & 1 & 0 & 0 & 0 & 0 & 0 & 0 \\
7 & 1 & 0 & 0 & 0 & 0 & 0 & 0 & 0 & 0 & 0 & 0 & 0 & 0 \\
8 & 0 & 0 & 0 & 0 & 1 & 0 & 0 & 0 & 1 & 0 & 0 & 0 & 0 \\
9 & 0 & 0 & 0 & 0 & 0 & 0 & 0 & 0 & 0 & 0 & 1 & 0 & 0 \\
10& 0 & 0 & 0 & 0 & 0 & 1 & 0 & 0 & 0 & 0 & 0 & 1 & 1 \\
11& 0 & 1 & 0 & 0 & 0 & 0 & 0 & 1 & 0 & 0 & 0 & 0 & 0 \\
12& 0 & 0 & 1 & 0 & 0 & 0 & 0 & 0 & 0 & 0 & 0 & 0 & 1 \\
13& 0 & 0 & 0 & 0 & 0 & 1 & 0 & 0 & 1 & 0 & 0 & 0 & 0 \\
14& 0 & 1 & 0 & 0 & 0 & 0 & 0 & 0 & 0 & 0 & 1 & 0 & 1 \\
15& 0 & 1 & 0 & 0 & 0 & 0 & 0 & 0 & 0 & 0 & 0 & 1 & 0 \\
16& 0 & 0 & 0 & 0 & 1 & 0 & 0 & 0 & 0 & 0 & 0 & 0 & 0 \\
17& 0 & 1 & 0 & 1 & 0 & 0 & 0 & 0 & 0 & 0 & 1 & 0 & 0 \\
18& 1 & 0 & 0 & 1 & 0 & 1 & 0 & 0 & 0 & 0 & 0 & 0 & 0 \\
19& 0 & 1 & 0 & 0 & 1 & 1 & 0 & 1 & 0 & 0 & 0 & 0 & 1 \\
20& 0 & 1 & 1 & 1 & 0 & 0 & 0 & 1 & 0 & 0 & 0 & 0 & 0 \\
21& 0 & 0 & 1 & 0 & 0 & 1 & 0 & 0 & 0 & 0 & 0 & 0 & 0 \\
22& 0 & 1 & 0 & 0 & 0 & 0 & 0 & 0 & 0 & 0 & 0 & 0 & 0 \\
23& 0 & 0 & 0 & 1 & 0 & 0 & 0 & 0 & 1 & 0 & 0 & 0 & 0 \\
\bottomrule
\end{tabular}%
}
\end{table}%

\clearpage

\begin{table}[!htbp]
\renewcommand\thetable{B4} 
\centering
\caption{The Q-matrix reported in \textcite{su_hierarchical_2013}}
\resizebox{\linewidth}{!}{
\begin{tabular}{cccccccccccccc}
\toprule
\multicolumn{14}{l}{Expert knowledge \parencite{su_hierarchical_2013}} \\
\midrule
\multirow{2}[4]{*}{Item} & \multicolumn{13}{c}{Attributes} \\
\cmidrule{2-14}& A1& A2& A3& A4& A5& A6& A7& A8& A9& A10 & A11 & A12 & A13 \\
\midrule
1 & 1 & 0 & 0 & 0 & 0 & 0 & 0 & 0 & 0 & 0 & 1 & 0 & 1 \\
2 & 0 & 0 & 0 & 0 & 0 & 1 & 0 & 0 & 0 & 0 & 0 & 0 & 0 \\
3 & 0 & 1 & 0 & 0 & 0 & 0 & 1 & 0 & 0 & 0 & 0 & 0 & 0 \\
4 & 0 & 0 & 0 & 1 & 0 & 0 & 0 & 0 & 1 & 0 & 0 & 0 & 0 \\
5 & 0 & 0 & 0 & 0 & 0 & 1 & 0 & 0 & 0 & 1 & 0 & 1 & 0 \\
6 & 0 & 0 & 0 & 0 & 0 & 1 & 1 & 0 & 0 & 0 & 0 & 0 & 0 \\
7 & 1 & 0 & 0 & 0 & 0 & 0 & 0 & 0 & 0 & 0 & 0 & 0 & 0 \\
8 & 0 & 0 & 0 & 0 & 1 & 0 & 0 & 0 & 1 & 0 & 0 & 0 & 0 \\
9 & 0 & 0 & 0 & 0 & 0 & 0 & 0 & 0 & 0 & 0 & 1 & 0 & 0 \\
10& 0 & 0 & 0 & 0 & 0 & 1 & 0 & 0 & 0 & 0 & 0 & 0 & 0 \\
11& 0 & 1 & 0 & 0 & 0 & 0 & 0 & 1 & 0 & 0 & 0 & 0 & 0 \\
12& 0 & 0 & 1 & 0 & 0 & 0 & 0 & 0 & 0 & 0 & 0 & 0 & 1 \\
13& 0 & 0 & 0 & 0 & 1 & 0 & 0 & 0 & 0 & 0 & 0 & 0 & 0 \\
14& 0 & 0 & 0 & 0 & 0 & 1 & 0 & 0 & 0 & 0 & 0 & 0 & 0 \\
15& 0 & 1 & 0 & 0 & 0 & 0 & 0 & 0 & 0 & 0 & 0 & 1 & 0 \\
16& 0 & 0 & 0 & 0 & 1 & 0 & 0 & 0 & 0 & 0 & 0 & 0 & 0 \\
17& 0 & 0 & 0 & 1 & 0 & 0 & 0 & 0 & 0 & 0 & 0 & 0 & 0 \\
18& 0 & 0 & 1 & 0 & 0 & 0 & 0 & 0 & 1 & 0 & 1 & 0 & 1 \\
19& 0 & 1 & 0 & 0 & 0 & 0 & 0 & 0 & 0 & 0 & 0 & 0 & 0 \\
20& 1 & 0 & 0 & 0 & 0 & 0 & 0 & 0 & 0 & 0 & 0 & 0 & 0 \\
21& 0 & 0 & 0 & 0 & 1 & 0 & 0 & 0 & 0 & 0 & 0 & 0 & 0 \\
22& 0 & 1 & 0 & 0 & 0 & 0 & 0 & 0 & 0 & 0 & 0 & 0 & 0 \\
23& 0 & 0 & 0 & 1 & 0 & 0 & 0 & 0 & 1 & 0 & 0 & 0 & 0 \\
\bottomrule
\end{tabular}%
}
\end{table}%

\clearpage

\subsection*{Supplementary Material C}
\subsubsection*{AIC and BIC under other DCMs}
AIC and BIC were calculated given the estimated Q-matrix from the proposed method. The TIMSS 2003 mathematics dataset was used for a comprehensive comparison among other DCMs. The results indicated that, although the proposed method assumes the DINA model, its estimated Q-matrix yielded a better fit than the Q-matrix by the domain experts in terms of AIC for DINA, DINO, RRUM, ACDM, and GDINA models and in terms of BIC for DINA, DINO, ACDM, and GDINA models.

\begin{table}[!htbp]
\renewcommand\thetable{C1} 
\centering
\caption{The values of AIC and BIC given the five DCMs}
\begin{center}
\scalebox{1}{
\begin{tabular}{rD{.}{.}{5} D{.}{.}{6} D{.}{.}{-1}}
\toprule
\multicolumn{4}{c}{AIC} \\
\midrule
\multicolumn{1}{r}{} & \multicolumn{1}{c}{Proposed method} & \multicolumn{1}{c}{Expert knowledge} & \multicolumn{1}{c}{Difference} \\
\midrule
DINA& 35420.71& 35743.76& -323.05\\
DINO& 35671.89& 35706.96& -35.07\\
RRUM& 34206.34& 34246.55& -40.21\\
ACDM& 33960.71& 34353.29& -392.58\\
GDINA & 34356.91& 34725.16& -368.25\\
\bottomrule
\end{tabular}%
}
\end{center}

\begin{center}
\scalebox{1}{
\begin{tabular}{rD{.}{.}{4} D{.}{.}{6} D{.}{.}{-1}}
\toprule
\multicolumn{4}{c}{BIC} \\
\midrule
\multicolumn{1}{r}{} & \multicolumn{1}{c}{Proposed method} & \multicolumn{1}{c}{Expert knowledge} & \multicolumn{1}{c}{Difference} \\
\midrule
DINA& 73552.78& 73875.83& -323.05\\
DINO& 73803.95& 73839.03& -35.07\\
RRUM& 72477.28& 72448.06& 29.23\\
ACDM& 72231.65& 72554.79& -323.14\\
GDINA & 72970.42& 73051.65& -81.23\\
\bottomrule
\end{tabular}
}
\end{center}
 \begin{tablenotes}
 \item \footnotesize{\textit{Note.} The differences in relative model fit indices were computed by subtracting the value of \emph{Expert knowledge} from that of \emph{Proposed method}.}
 \end{tablenotes}
\end{table}

\clearpage
\color{black}

\subsection*{Supplementary Material D}
\subsubsection*{Comparison between the proposed method and other Bayesian methods with identification constraints}

To investigate the possibility in which the proposed method performs poorly compared with other Bayesian methods with identification constraints in situations where a true Q-matrix possesses a certain design yielding identified model parameters, we conducted an additional simulation study based on the specification adopted in \textcite{liu_constrained_2020}. Specifically, we considered the conditions of $K=3$ or 4, $N=500$, $J=18$, and $\rho=0$ or 0.25. The same attribute-pattern generation method as \textcite{liu_constrained_2020} was employed in this simulation study. The true Q-matrices were also specified identical to \textcite{liu_constrained_2020}. Further, the same setting of the proposed method---as in the condition of $N=500$ in our simulation study---was employed for our estimation.
Table \ref{tab:comparewithliu} presents the results of matrix- and element-wise recovery rates. They show that the recovery rates from the proposed method were highly comparable with those from \textcite{liu_constrained_2020}, despite the fact that our method estimated a Q-matrix without the knowledge of whether a true Q-matrix in the simulation condition holds an identified structure. This is in contrast to the methods in \textcite{liu_constrained_2020} in the sense that they estimated a Q-matrix using the identification constraints when the true Q-matrix in the simulation conditions is known to have such a structure. Therefore, these results suggest that our method estimated an identified, true Q-matrix without any identification constraints in an accurate and fast manner from a much greater Q-matrix space than the space being considered by the methods with the identification constraints in \textcite{liu_constrained_2020}.

\begin{table}[!htbp]
\color{black}
\renewcommand\thetable{D1} 
\caption{Recovery rate under the simulation conditions specified in \textcite{liu_constrained_2020}.}
\resizebox{\linewidth}{!}{
\begin{tabular}{ccccccccccccc}
\toprule
& && & \multicolumn{4}{c}{matrix-wise recovery (\%)} & & \multicolumn{4}{c}{element-wise recovery (\%)} \\
\cmidrule{5-8}\cmidrule{10-13} $K$ & $N$&$\rho$ &$J$ & cGibbs1 & cGibbs20 & cMHRM & \begin{tabular}{c}Proposed \\ Method \end{tabular}& & cGibbs1 & cGibbs20 & cMHRM & \begin{tabular}{c}Proposed \\ Method \end{tabular} \\
\midrule
3 & 500 & 0 & 18& 99& 100 & 100 & 100& & 99.6& 100 & 100 &100\\
3 & 500 & 0.25& 18& 100 & 100 & 100 &100 & & 100 & 100 & 100 &100\\
4 & 500 & 0 & 18& 96& 98& 98&98 & & 99.4& 99.96 & 99.96 &99.72\\
4 & 500 & 0.25& 18& 98& 99& 99&98 & & 99.7& 99.99 & 99.99 & 99.72 \\
\bottomrule
\end{tabular}%
}
 \begin{tablenotes}
\footnotesize \item \textit{Note}. The recovery rates of cGibbs1, cGibbs20, and cMHRM were adapted directly from \textcite{liu_constrained_2020}.
 \end{tablenotes}
\label{tab:comparewithliu}%
\end{table}

\clearpage

\subsection*{Supplementary Material E}
We considered three cases where the log-likelihood value of an unidentified, true Q-matrix equals that of another Q-matrix, whereas the ELBO value of the true one differs from that of another one. 

The following settings were assigned to the three cases. The sample size was set to $N=100,000$. The true values of the guessing and slip parameters for the DINA model were specified as in Table \ref{tab:Supp_E_truevalues}. To generate datasets, we used the \texttt{simGDINA} function in the \texttt{GDINA} R package \parencite{gdina_R_2020}. The stopping criteria for the proposed algorithm and marginal likelihood estimation with an EM algorithm were specified such that the iteration stopped when the maximum change in the ELBO or two-times the negative log likelihood became less than $10^{-6}$ or the number of iterations reached 3000. Lastly, non-informative prior distributions were assigned to structural and item parameters, and initial values of latent indicator variable $z_{il}$ were set to be $1/L$ for the proposed method. 

% Table generated by Excel2LaTeX from sheet 'K=2'
\begin{table}[!htbp]
\renewcommand\thetable{E1} 
\centering
\caption{Specification of guessing and slip parameters}
\begin{tabular}{ccc}
\toprule
& \multicolumn{2}{c}{True Item Parameters} \\
\midrule
Item& Guessing & Slip \\
\midrule
1 & 0.290 & 0.227 \\
2 & 0.262 & 0.236 \\
3 & 0.057 & 0.100 \\
4 & 0.231 & 0.169 \\
5 & 0.239 & 0.096 \\
6 & 0.073 & 0.294 \\
\bottomrule
\end{tabular}%
\label{tab:Supp_E_truevalues}%
\end{table}%

\clearpage
\subsubsection*{Example 1: $K=2$ and $J=6$}
Tables \ref{tab:E2} and \ref{tab:E3} show the details of Q-matrices in this example and their values of log likelihood and ELBO. Although the log-likelihood value of the true Q-matrix equals that of another one, the ELBO correctly prefers the true Q-matrix over another one. 

\begin{table}[!htbp]
\renewcommand\thetable{E2} 
\centering
\caption{Q-matrices for the case with $K=2$ and $J=6$}
\resizebox{!}{!}{
\begin{tabular}{ccccccc}
\toprule
\multicolumn{7}{c}{$K=2$} \\
\midrule
True Q-matrix & \multicolumn{2}{c}{Attribute} & & Another Q-matrix & \multicolumn{2}{c}{Attribute} \\
\cmidrule{1-3}\cmidrule{5-7}Item& 1 & 2 & & Item& 1 & 2 \\
\cmidrule{1-3}\cmidrule{5-7}1 & 1 & 0 & & 1 & 1 & 0 \\
2 & 1 & 0 & & 2 & 1 & 0 \\
3 & 1 & 0 & & 3 & 1 & 0 \\
4 & 1 & 1 & & 4 & 1 & 1 \\
5 & \cellcolor{red}{1}
& 1 & & 5 & \cellcolor{red}{0} & 1 \\
6 & \cellcolor{red}{1} & 1 & & 6 & \cellcolor{red}{0} & 1 \\
\bottomrule
\end{tabular}%
}
 \begin{tablenotes}
\footnotesize \item \textit{Note}. The red cell boxes denote the entries that changed from the true Q-matrix to another one.
 \end{tablenotes}
\label{tab:E2}%
\end{table}%

\begin{table}[!htbp]
\renewcommand\thetable{E3} 
\centering
\caption{Values of log likelihood and ELBO}
\begin{tabular}{ccc}
\toprule
& Log Likelihood& ELBO \\
\midrule
True Q-matrix & -320046.490 & -320125.019 \\
Another Q-matrix & -320046.490 & -320128.646 \\
\bottomrule
\end{tabular}%
\label{tab:E3}%
\end{table}%

\clearpage
\subsubsection*{Example 2: $K=3$ and $J=6$}
Tables \ref{tab:E4} and \ref{tab:E5} show the details of Q-matrices in this example and their values of log likelihood and ELBO. Although the log-likelihood value of the true Q-matrix equals that of another one, the ELBO correctly prefers the true Q-matrix over another one. 

\begin{table}[!htbp]
\renewcommand\thetable{E4} 
\centering
\caption{Q-matrices for the case with $K=3$ and $J=6$}
\begin{tabular}{ccccccccc}
\toprule
\multicolumn{9}{c}{K=3} \\
\midrule
True Q-matrix & \multicolumn{3}{c}{Attribute} & & Another Q-matrix & \multicolumn{3}{c}{Attribute} \\
\cmidrule{1-4}\cmidrule{6-9}Item& 1 & 2 & 3 & & Item& 1 & 2 & 3 \\
\cmidrule{1-4}\cmidrule{6-9}1 & 1 & 0 & 0 & & 1 & 1 & 0 & 0 \\
2 & 0 & 0 & 1 & & 2 & 0 & 0 & 1 \\
3 & 1 & 0 & 1 & & 3 & 1 & 0 & 1 \\
4 & 1 & 0 & 1 & & 4 & 1 & 0 & 1 \\
5 & 1 & 1 & \cellcolor{red}{1} & & 5 & 1 & 1 & \cellcolor{red}{0} \\
6 & 1 & 1 & 1 & & 6 & 1 & 1 & 1 \\
\bottomrule
\end{tabular}%
 \begin{tablenotes}
\footnotesize \item \textit{Note}. The red cell box denotes the entries that changed from the true Q-matrix to another one.
 \end{tablenotes}
\label{tab:E4}%
\end{table}%

\begin{table}[!htbp]
\renewcommand\thetable{E5} 
\centering
\caption{Values of log likelihood and ELBO}
\begin{tabular}{ccc}
\toprule
& Log Likelihood& ELBO \\
\midrule
True Q-matrix & -347569.953 & -347668.958 \\
Another Q-matrix & -347569.953 & -347671.755 \\
\bottomrule
\end{tabular}%
\label{tab:E5}%
\end{table}%

\clearpage
\subsubsection*{Example 3: $K=4$ and $J=6$}
Tables \ref{tab:E6} and \ref{tab:E7} show the details of Q-matrices in this example and their values of log likelihood and ELBO. Although the log-likelihood value of the true Q-matrix equals that of another one, the ELBO correctly prefers the true Q-matrix over another one. 

\begin{table}[!htbp]
\renewcommand\thetable{E6} 
\centering
\caption{Q-matrices for the case with $K=4$ and $J=6$}
\begin{tabular}{ccccccccccc}
\toprule
\multicolumn{11}{c}{K=4} \\
\midrule
True Q-matrix & \multicolumn{4}{c}{Attribute} & & Another Q-matrix & \multicolumn{4}{c}{Attribute} \\
\cmidrule{1-5}\cmidrule{7-11}Item& 1 & 2 & 3 & 4 & & Item& 1 & 2 & 3 & 4 \\
\cmidrule{1-5}\cmidrule{7-11}1 & 1 & 0 & 0 & 0 & & 1 & 1 & 0 & 0 & 0 \\
2 & 1 & 1 & 0 & 0 & & 2 & 1 & 1 & 0 & 0 \\
3 & 1 & 0 & 1 & 0 & & 3 & 1 & 0 & 1 & 0 \\
4 & 1 & 1 & 1 & 0 & & 4 & 1 & 1 & 1 & 0 \\
5 & 1 & 1 & \cellcolor{red}{1} & 1 & & 5 & 1 & 1 & \cellcolor{red}{0} & 1 \\
6 & 1 & 1 & \cellcolor{red}{1} & 1 & & 6 & 1 & 1 & \cellcolor{red}{0} & 1 \\
\bottomrule
\end{tabular}%
 \begin{tablenotes}
\footnotesize \item \textit{Note}. The red cell boxes denote the entries that changed from the true Q-matrix to another one.
 \end{tablenotes}
\label{tab:E6}%
\end{table}%

\begin{table}[!htbp]
\renewcommand\thetable{E7} 
\centering
\caption{Values of log likelihood and ELBO}
\begin{tabular}{ccc}
\toprule
& Log Likelihood& ELBO \\
\midrule
True Q-matrix & -346630.739 & -346763.324 \\
Another Q-matrix & -346630.739 & -346765.463 \\
\bottomrule
\end{tabular}%
\label{tab:E7}%
\end{table}%

\end{document}